\documentclass[draftclsnofoot,10pt,onecolumn]{IEEEtran}

\usepackage{amssymb, amsthm, graphicx, cite, url}
\usepackage[caption=false,font=footnotesize]{subfig}
\usepackage[cmex10]{amsmath}

\newtheorem{thm}{Theorem}
\newtheorem{cor}{Corollary}
\newtheorem{lem}{Lemma}
\newtheorem{prop}{Proposition}
\newtheorem{defi}{Definition}

\newcommand{\wh}[1]{\widehat{#1}}
\newcommand{\wt}[1]{\widetilde{#1}}
\newcommand{\maxx}{{\sf max}}
\newcommand{\Rsum}{R_{\sf sum}}
\newcommand{\LLR}{{\sf LLR}}

\def\e{\epsilon}
\def\d{\delta}
\newcommand{\Ec}{\mathcal{E}}
\newcommand{\Xc}{\mathcal{X}}
\newcommand{\Kh}{{\widehat{K}}}
\newcommand{\kh}{{\widehat{k}}}
\newcommand{\Rh}{{\widehat{R}}}
\newcommand{\aep}{{\mathcal{T}_{\epsilon}^{(n)}}}
\newcommand{\aepvar}{{\mathcal{T}_{\epsilon'}^{(n)}}}
\DeclareMathOperator*{\argmin}{\arg\min}
\DeclareMathOperator*{\argmax}{\arg\max}

\title{A Context-Aware CEO Problem}

\author{ Daewon Seo, Sung Hoon Lim, and Yongjune Kim 
\thanks{ This work was supported in part by the National Research Foundation of Korea (NRF) Grant funded by the Korea Government (MSIT, Ministry of Science and ICT) under Grant RS-2023-00253008, NRF-2020R1F1A1074926, and RS-2023-00212103. }
\thanks{ D.~Seo is with the Department of Electrical Engineering and Computer Science, Daegu Gyeongbuk Institute of Science and Technology (DGIST), Daegu 42988, South Korea (e-mail: dwseo@dgist.ac.kr). S.~H.~Lim is with the School of Information Sciences, Hallym University, Chuncheon 24252, South Korea (e-mail: shlim@hallym.ac.kr). Y.~Kim is with the Department of Electrical Engineering, Pohang University of Science and Technology (POSTECH), Pohang 37673, South Korea, and also with the Institute for Convergence Research and Education in Advanced Technology, Yonsei University, Seoul 03722, South Korea (e-mail: yongjune@postech.ac.kr). }
}

\begin{document}
\maketitle

\begin{abstract}
In many sensor network applications, a fusion center often has additional valuable information, such as context data, which cannot be obtained directly from the sensors. Motivated by this, we study a generalized CEO problem where a CEO has access to context information. The main contribution of this work is twofold. Firstly, we characterize the asymptotically optimal error exponent per rate as the number of sensors and sum rate grow without bound. The proof extends the Berger-Tung coding scheme and the converse argument by Berger \textit{et al}.~(1996) taking into account context information. The resulting expression includes the minimum Chernoff divergence over context information. Secondly, assuming that the sizes of the source and context alphabets are respectively $|\mathcal{X}|$ and $|\mathcal{S}|$, we prove that it is asymptotically optimal to partition all sensors into at most $\binom{|\mathcal{X}|}{2} |\mathcal{S}|$ groups and have the sensors in each group adopt the same encoding scheme. Our problem subsumes the original CEO problem by Berger \textit{et al}.~(1996) as a special case if there is only one letter for context information; in this case, our result tightens its required number of groups from $\binom{|\mathcal{X}|}{2}+2$ to $\binom{|\mathcal{X}|}{2}$. We also numerically demonstrate the effect of context information for a simple Gaussian scenario.
\end{abstract}


\section{Introduction} \label{sec:intro}
In the era of sensor networks, the Internet of Things (IoT), and many emerging applications such as autonomous vehicles, numerous sensors or devices collect data, from which a fusion center performs a specific task. A common bottleneck in such scenarios is the communication resources between the sensors and the fusion center. For instance, sensors in an autonomous vehicle generate up to 25 gigabytes (GB) of data every hour \cite{McKinsey2014}. Hence, it is often essential for the sensors to compress or quantize the collected data before delivering it to the fusion center. Additionally, in ensemble learning, many weak classifiers make initial inferences from data, and then a strong classifier (i.e., the fusion center) makes a final inference~\cite{Freund1997,Kim2023distributed}. The initial inferences can be thought of as compressed versions of noisy data; thus, it raises questions about how to compress noisy data and what the optimal error probability is.

Another critical aspect of such applications is that a resource-abundant fusion center often has other information relevant to the task that local sensors cannot provide. For instance, in object recognition, multiple cameras take photos of the scene and transmit compressed versions to the fusion center. The quality of photos for the task depends on other factors such as exposure, lighting conditions, distance and angle from the object, etc., which we call context information.\footnote{It is also called side information in information theory literature.} The fusion center cannot accurately perceive the context information from the pictures alone; it usually equips with an extra sensor dedicated to the context information. Then, an interesting question arises as to how helpful such context information is for the task.

To answer the above questions, we consider the CEO problem by Berger \textit{et al}.~\cite{BergerZV1996} and additionally introduce context information available to the CEO. Specifically, the CEO of an organization aims to detect a length-$n$ sequence of discrete random variables $X^n \in \mathcal{X}^n$. However, instead of observing $X^n$ directly, $L$ local sensors collect noisy observations of $X^n$ through identical observational channels. These sensors report their observations to the CEO via rate-constrained noiseless links. Unlike canonical CEO problems, context information in $Y^n$ is assumed to be available to the CEO via a separate link. Upon receiving compressed observations and context information, the CEO infers the unknown $X^n$ as accurately as possible.

The original CEO problem that has no context information is first proposed by Berger \textit{et al}.~\cite{BergerZV1996}, which characterizes the asymptotically optimal error exponent per rate as $L$ tends to infinity. To attain it, the authors propose to partition sensors into at most $\binom{|\mathcal{X}|}{2}+2$ groups, and the sensors in each group have the same encoding scheme. The original CEO problem is extended to the quadratic Gaussian CEO problem in \cite{ViswanathanB1997, Oohama1998}, where the asymptotically optimal tradeoff between sum-rate and mean-squared error (MSE) is characterized. Relying on the properties of the Gaussian distribution, the exact rate region for the quadratic Gaussian CEO problem having a finite number of sensors is characterized \cite{Oohama2005, PrabhakaranTR2004}. Another well-understood CEO problem is with logarithmic distortion, for which the exact rate region for a general setting is given \cite{CourtadeW2014}. Several extensions have been further studied, such as Byzantine agents \cite{KosutT2009}, multiple sources \cite{YangX2012}, vector Gaussian \cite{EkremU2014, RiniKSG2019, UgurAZ2020} and non-Gaussian \cite{VempatyV2015b, SeoV2021} problems, specific coding structures \cite{NangirACAM2019, BehrooziS2009}, and binary source \cite{HeZKJM2016}. The most renowned theoretical understanding of data compression in the presence of context information is by Wyner and Ziv \cite{WynerZ1976}, which assumes that the context information is available to the decoder. This approach is extended to various problems, such as multiterminal source coding \cite{DiggaviV2004} and remote source coding \cite{YamamotoI1980}. However, to the best of our knowledge, it has not yet been studied in the setting of CEO problems.

This work, in particular, focuses on the asymptotic property of the CEO problem where the number of sensors and sum rate grow without bound, as in \cite{BergerZV1996, ViswanathanB1997, Oohama1998, KosutT2009}. In this case, as we will see, the problem bears some similarity with distributed detection \cite{Varshney1997, VeeravalliBP1993, SaligramaAS2006, HellmanC1970, TayTW2008}, which studies the area from a perspective of hypothesis testing: Instead of (possibly infinitely long) block compression of the CEO problem, local sensors make decisions in a symbol-wise manner that can be thought of as symbol-wise compression. In the same model as the original CEO problem, but with only symbol-wise compression being allowed, Tsitsiklis \cite{Tsitsiklis1988} characterized the asymptotically optimal error exponent when the number of sensors grows without bound. It also shows that to achieve the optimal error exponent, it is sufficient for the sensors to be partitioned into at most $\binom{|\mathcal{X}|}{2}$ groups and for all sensors in each group to perform the same scalar compression. Then, recalling $\binom{|\mathcal{X}|}{2}+2$ of the original CEO problem, it is natural to ask whether or not such +2 of the CEO problem can be further tightened. By the argument of linear fractional programming (LFP), we conclude that having at most $\binom{|\mathcal{X}|}{2}$ groups is indeed sufficient for the CEO problem without context information as well.

This work extends and integrates the aforementioned works to the case where the CEO utilizes context information for inference. The main contributions of this work can be summarized as follows.
\begin{itemize}
	\item We consider the CEO problem where the CEO has context information and characterize its asymptotically optimal error exponent per rate (Theorem \ref{thm:alpha}). The expression is written in terms of mutual information and the minimum Chernoff divergence over context and source alphabets, which extends the existing characterization by Berger \textit{et al}.~\cite{BergerZV1996}. The achievability is based on the Berger-Tung scheme with additional consideration of context information and the error probability analysis of hypothesis testing. The converse is based on the converse arguments of coding rate and the error probability of hypothesis testing.
	
	\item To attain the optimal error exponent per rate, we prove that it is sufficient to divide all sensors into at most $\binom{|\mathcal{X}|}{2} |\mathcal{S}|$ groups, where $|\mathcal{X}|$ and $|\mathcal{S}|$ are the sizes of the source and context alphabets, and all sensors in each group adopt the same compression scheme (Theorem \ref{thm:reduced_cardinality}). When $|\mathcal{S}|=1$, it tightens Berger \textit{et al.}'s number of groups $\binom{|\mathcal{X}|}{2}+2$ for the CEO problem and coincides with Tsitsiklis's number. The same proof technique can also be applied to another CEO problem \cite[Theorem 2]{BergerZV1996}, which derives $\binom{|\mathcal{X}|}{2}$ bound, and other detection problems \cite{Tsitsiklis1988, NitinawaratAV2013}. The proof relies on the argument of linear fractional programming (LFP).
\end{itemize}

The rest of this paper is organized as follows. Section \ref{sec:formulation} formally defines the problem of interest, presents the main theorem, and discusses its implications. Sections \ref{sec:achievability}, \ref{sec:converse}, and \ref{sec:cardinality} respectively provide the proofs of achievability, converse, and the number of groups required to attain the optimal exponent per rate. Section \ref{sec:example} numerically demonstrates the gain of context information for a simple Gaussian scenario. Finally, Section \ref{sec:conclusion} concludes the paper.

\section{Problem Statement and Main Result} \label{sec:formulation}
\subsection{Problem Statement} \label{subsec:problem_statement}
\begin{figure}[t]
	\centering
	\includegraphics[width=3.0in]{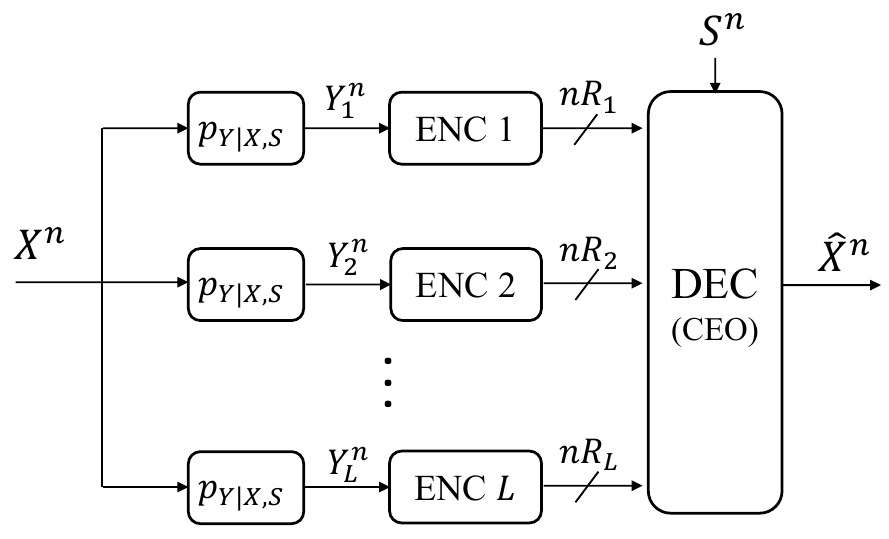}
	\caption{The problem model with $L$ sensors. If the context information is fixed to be a certain element, i.e., $|\mathcal{S}| = 1$, it reduces to the original CEO problem \cite{BergerZV1996}. If $|\mathcal{S}| = 1$ and only scalar compression at a fixed rate per sensor is allowed, i.e., $n=1$ and $R_\ell=\text{constant}$ for all $\ell$, then it reduces to the distributed detection problem in \cite{Tsitsiklis1988}.}
	\label{fig:model}
\end{figure}

We consider the CEO problem with discrete alphabets, depicted in Figure \ref{fig:model}. Suppose that there is a probability mass function (pmf) $p_{X,S}(x,s)$ over $\mathcal{X} \times \mathcal{S}$ and a sequence of source and context information pair $(X^n, S^n)$ where $X^n = (X(1), \ldots, X(t), \ldots, X(n)) \in \mathcal{X}^n$ and $S^n = (S(1), \ldots, S(t), \ldots, S(n)) \in \mathcal{S}^n$ that is independent and identically distributed (i.i.d.) drawn from $p_{X,S}(x,s)$. We further suppose that $p_X(x) = \sum_{s \in \mathcal{S}} p_{X,S}(x,s) > 0$ for all $x \in \mathcal{X}$ and $p_S(s) = \sum_{x \in \mathcal{X}} p_{X,S}(x,s) > 0$ for all $s \in \mathcal{S}$ because removing $x$ with $p_X(x) = 0$ and $s$ with $p_S(s) = 0$ does not affect the problem. Also, there are $L$ sensors, each of which measures noisy data from the source. The $\ell$-th sensor observes $Y_\ell^n \in \mathcal{Y}^n$ according to a common observational channel law $p_{Y|X, S}(y_\ell(t)|x(t), s(t))$. We assume that $p_{Y|X,S}(y|x,s) \ne p_{Y|X,S}(y|x',s')$ if $(x,s) \ne (x',s')$; otherwise the source is indistinguishable for those $(x,s)$ and $(x',s')$ such that $p_{Y|X,S}(y|x,s) = p_{Y|X,S}(y|x',s')$. For the fusion center, commonly referred to as a CEO, to be able to identify the source, each sensor encodes $Y_{\ell}^n$ into a codeword $C_\ell \in \mathcal{C}_\ell$ of rate $R_\ell$ nats. Sensors transmit codewords, or more precisely, the indices of codewords, subject to the total rate constraint $\sum_{\ell=1}^{L} R_\ell \le \Rsum$, to the CEO. In addition to the received codewords, the CEO is aware of context information $S^n \in \mathcal{S}^n$ drawn jointly with $X^n$ from $p_{X,S}(x,s)$.

Upon receiving context information $S^n$ and $L$ codeword indices $C^{L} = \{C_\ell\}_{\ell=1}^L$ from sensors, the CEO makes estimates $\wh{X}^n(C^L, S^n) = \{\wh{X}(t)\}_{t=1}^n$ that minimizes estimation error frequency
\begin{align*}
	P_e^{(n)} = \frac{1}{n} \mathbb{E}\left[ d_H( X^n, \wh{X}^n )\right] = \frac{1}{n} \sum_{t=1}^n \mathbb{E} \left[ d_H( X(t), \wh{X}(t) )\right],
\end{align*}
where $d_H$ is the Hamming distortion.

This setting is a variant of the original \emph{CEO problem} by Berger \textit{et al}.~\cite{BergerZV1996}, as which we focus on the asymptotic tradeoff between the sum rate and the error probability as $L , \Rsum \to \infty$. Formally, define two quantities:
\begin{align*}
	P_e^{(n)}(L, \Rsum) &= \min_{\sum_{\ell=1}^L R_\ell \le \Rsum} P_e^{(n)}(C_1, \ldots, C_L, S^n), \\
	P_e(\Rsum) &= \lim_{L \to \infty} P_e(L, \Rsum) \\
	&= \lim_{L \to \infty} \lim_{n \to \infty} P_e^{(n)} (L, \Rsum).
\end{align*}
As one can see, $P_e^{(n)}(L, \Rsum)$ is the smallest error probability that is achievable using the best codebook of block length $n$ and sum rate $\Rsum$ with $L$ sensors. Allowing an infinitely long block coding and sending $L \to \infty$, $P_e(\Rsum)$ indicates the smallest error probability that is achievable at $\Rsum$, or in other words, $-\log P_e(\Rsum)$ is the largest achievable error exponent.
Then, as in \cite{BergerZV1996} the goal of this work is to characterize
\begin{align}
	\alpha = \alpha(p_{X,S}, p_{Y|X,S}) := \lim_{\Rsum \to \infty} \frac{-\log P_e(\Rsum)}{\Rsum}, \label{eq:alpha_def}
\end{align}
i.e., $\alpha$ is the largest error exponent per nat (or bit) when a large sum rate is provided. Note that the inner expression of \eqref{eq:alpha_def} is with respect to a fixed sum rate, but the number of sensors has already diverged. Hence, our interest is a regime where the average rate per sensor vanishes slowly with $L$ so that $\Rsum$ diverges. It should be also noted that when $|\mathcal{S}|=1$, it reduces to the original CEO problem \cite{BergerZV1996}. Hence, this work aims to characterize $\alpha$ when $|\mathcal{S}| \ge 2$.

\medskip
\noindent\emph{Notation:} Throughout the paper, $\ell$ and $t$ denote the sensor index and temporal index of block coding, respectively. Also, to simplify notation, we consider a set with $k$ element across $\ell$ and/or $t$ as a vector of length $k$. For instance, $Y_\ell^{n} = \{Y_\ell(t)\}_{t=1}^n$, $Y^{L}(t) = \{Y_\ell(t)\}_{\ell=1}^L$, and $Y^{nL} = \{Y_\ell^n\}_{\ell=1}^L$. If the $t$-th element of a length $n$ vector is omitted, we use temporal index $t^c$; for instance, $X(t^c) = \{X(1), \ldots, X(t-1), X(t+1), \ldots, X(n)\}$. A set of consecutive integers are represented by $[i:j] := \{i, i+1, \ldots, j\}$. Also, a subscript of probability mass functions is often omitted when it is clear from context; for instance, $p(x) = p_X(x), p(y_\ell(t)|x(t)) = p_{Y|X}(y_\ell(t)|x(t))$. All logarithms are natural logarithms; thus, the unit of all information-theoretic quantities is nats.

\subsection{Main Result} \label{subsec:main_result}
Before formally discussing the main result, we introduce the Chernoff divergence, one of the key functionals of this work.
\begin{defi}
	Between two probability distributions $p_0$ and $p_1$ on $\mathcal{Z}$ that are dominated by $\mu$, the Chernoff divergence $d_\lambda$ of parameter $\lambda \in [0,1]$ is defined as
	\begin{align*}
		d_\lambda(p_0, p_1) := -\log \int_{\mathcal{Z}} p_0^{1-\lambda}(z) p_1^{\lambda}(z) d\mu(z).
	\end{align*}
	If $p_0$ and $p_1$ do not have a common support, we define $d_\lambda(p_0, p_1) = \infty$.
\end{defi}
The maximum value of the Chernoff divergence over $\lambda$ is in particular called the Chernoff information $C(p_0, p_1)$, i.e.,
\begin{align*}
	C(p_0, p_1) &:= \max_{\lambda \in [0,1]} d_\lambda(p_0, p_1) = -\log \min_{\lambda \in [0,1]} \int_{\mathcal{Z}} p_0^{1-\lambda}(z) p_1^{\lambda}(z) d\mu(z),
\end{align*}
which is widely used as it upper bounds the error exponent in the Bayesian binary hypothesis testing and becomes tight as the number of i.i.d.~observations tends to infinity \cite{Poor1988, MoulinV2019, CoverT1991}.

The optimal error exponent per rate of our CEO problem is as follows.
\begin{thm} \label{thm:alpha}
	Let $J$ be an independent auxiliary random variable on space $\mathcal{J}$ such that $|\mathcal{J}| = \binom{|\mathcal{X}|}{2}|\mathcal{S}|$, and let $p(u|y,j)$ be any probability mass function on $\mathcal{U}$ such that $|\mathcal{U}| = \left(\binom{|\mathcal{X}|}{2}|\mathcal{S}| + |\mathcal{Y}| \right) |\mathcal{J}|$. Then,
	\begin{align}
		\alpha &= \lim_{c \to 0} \max_{ \substack{p_J, p_{U|Y,J}: \\ I(U;Y|X,S,J)=c } } \frac{ \min_{(s, x_1, x_2)} \max_{\lambda \in [0,1]} \mathbb{E}_{J} \left[ d_{\lambda}(p_{x_1, s, J}(u), p_{x_2, s, J}(u)) \right] }{I(U;Y|X,S,J)}, \label{eq:alpha}
	\end{align}
	where
	\begin{align*}
		p_{x, s, j}(u) := p_{U|X,S,J}(u|x,s,j)  = \sum_y p(y|x,s)p(u|y,j).
	\end{align*}
\end{thm}
\begin{IEEEproof}
	The proofs of achievability and converse are respectively given in Sections \ref{sec:achievability} and \ref{sec:converse}. The proof of $|\mathcal{J}|$ is given in Section \ref{sec:cardinality}.
\end{IEEEproof}

In the expression, the auxiliary random variable $J$ plays the role of a group index, and $p_J$ denotes the fraction of sensors belonging to the $j$-th group that uses the same encoding scheme. Then, the outer maximum can be interpreted as a maximum over all possible fractions and coding schemes (test channels) under constraint. Fixing a particular fraction $p_J$ and a coding scheme for the $j$-th group $p_{U|Y,J=j}$, $I(U;Y|X,S,J=j)$ is an individual coding rate for a sensor in the $j$-th group; thus, the denominator is simply the individual coding rate averaged over all sensors since $I(U;Y|X,S,J) = \sum_j p_J(j)I(U;Y|X,S,J=j)$. In the end, the mutual information is sent to zero by $\lim_{c \to 0}$ operation for the regime of vanishing average rate per sensor. Before discussing the numerator, consider a symbolwise compression problem (i.e., no block coding) to get an insight, and in such a case, the probability of error can be decomposed as follows.
\begin{align*}
	P_e &= \sum_s p(s) \sum_{x} p(x|s) \sum_{x' \ne x} \mathbb{P}[ \wh{X}=x' |X=x, S=s ] \\
	&= \sum_s p(s) \sum_{(x, x'): x \ne x'} p(x|s) \mathbb{P}[ \wh{X}=x' |X=x, S=s ].
\end{align*}
Since the inner sum comprises $\binom{|\mathcal{X}|}{2}$ pairs of binary hypothesis testing's error events for a fixed $s$, and each error term $\mathbb{P}[ \wh{X}=x' |X=x, S=s ]$ decays exponentially fast when there are $L$ sensors and $L$ tends to infinity, the smallest error exponent dominates $P_e$. It means, maximizing the smallest exponent among individual error terms, found by $\min \max$ operation, optimizes the entire error exponent. In other words, the outer $\max$ operator maximizes an error exponent of the binary hypothesis testing for the most indistinguishable triplet of source and context information realizations. In addition, the CEO observes codewords and context information. Then, conditioned on $X=x$, the CEO's observational channel is effectively $p_{U^L, S | X} = p_{S|X} \prod_{\ell} p_{U_\ell | X, S}$. Since $\prod_{\ell} p_{U_\ell | X, S}$ dominates the observational channel as $L$ grows without bound, we have the sum of individual divergences across sensors using the additive property of the Chernoff divergence for product distributions. Converting the sum into the ensemble sum yields the numerator in its expectation form, i.e., $\mathbb{E}_{J} \left[ d_{\lambda}(p(u|x_1, s, J), p(u|x_2, s, J)) \right]$.

Note that the cardinality of $J$ in the theorem is $\binom{|\mathcal{X}|}{2} |\mathcal{S}|$. If the original CEO problem \cite{BergerZV1996} is of interest, i.e., $|\mathcal{S}|=1$, our setting reduces to $\binom{|\mathcal{X}|}{2}$, which means that sensors should be partitioned into at most $\binom{|\mathcal{X}|}{2}$ groups and each of which, informally speaking, corresponds to distinguishing each pair of hypotheses. This bound tightens the original cardinality bound $\binom{|\mathcal{X}|}{2}+2$ by Berger \textit{et al}.~\cite[Theorem 1]{BergerZV1996}. The $+2$ is from the support lemma \cite[Lemma 3.3.4]{CsiszarK1997}, \cite[Appendix C]{ElGamalK2011} that technically bounds the cardinality of auxiliary random variables. Our proof is based on the linear fractional programming (LFP) that subsumes the linear programming as a special case and is generally applicable to the CEO problem if the number of sensors and sum rate grow without bound. For instance, applying our LFP-based proof removes the additional $+2$ term in the expression for the CEO problem with nonexchangeable sensors \cite[Theorem 2]{BergerZV1996}.

As our problem and result bear several similarities to the works by Tsitsiklis \cite{Tsitsiklis1988} and the original CEO problem by Berger \textit{et al}.~\cite{BergerZV1996}, we highlight connections and differences here.
\begin{itemize}
	\item If $|\mathcal{S}|=1$ and only scalar compression is allowed, i.e., $n=1$, then the problem becomes a distributed detection problem in \cite{Tsitsiklis1988}, for which the asymptotically optimal error exponent in $L$ is as follows.
	\begin{align*}
		\wt{\alpha} = \max_{ p_J, p_{U|Y,J} } \min_{(x_1, x_2)} \max_{\lambda \in [0,1]} \mathbb{E}_{J} \left[ d_{\lambda}(p(u|x_1, J), p(u|x_2, J)) \right],
	\end{align*}
	where $\mathcal{U}$ is a finite space of compressed signals determined by a fixed compression rate and $p_{U|Y,J}$ is a (usually deterministic in scalar compression literature) mapping for compression. Note that the expression is the same as $\alpha$ ignoring the mutual information term for normalization. Moreover, $\wt{\alpha}$ can be attained by partitioning sensors into at most $|\mathcal{J}| = \binom{|\mathcal{X}|}{2}$ groups and sensors in each group adopt the same scalar compression scheme, i.e., sensors in the $j$-th group uses the same $p(u|y,j)$ mapping. The number of groups is the same as ours if $|\mathcal{S}|=1$.
	
	\item If $|\mathcal{S}|=1$, then the problem reduces to the original CEO problem by Berger \textit{et al.}~\cite{BergerZV1996}, which studies the optimal error exponent per rate in the regime of vanishing average rate per sensor. However, our expression \eqref{eq:alpha} is distinct from \cite[Theorem 1]{BergerZV1996} even when $|\mathcal{S}|=1$; ours includes additional mutual information control represented by $\lim_{I(\cdots) \to 0}$. This distinction arises because the expression in \cite[Theorem 1]{BergerZV1996} is over an unrestricted search space of test channels, allowing for an arbitrary average rate per sensor, which is outside the regime of vanishing average rate per sensor. In this sense, the original CEO problem \cite[Theorem 1]{BergerZV1996} has an inconsistency in the resulting expression and its problem formulation, which we have refined as in \eqref{eq:alpha} in this work. Also, Berger \textit{et al.}~characterized $\alpha$ with $|\mathcal{J}| = \binom{|\mathcal{X}|}{2} + 2$ groups \cite{BergerZV1996}. We have tightened it to $\binom{|\mathcal{X}|}{2}$.
\end{itemize}
The remaining sections are devoted to the proof of the theorem and its numerical demonstration.

\section{Proof of Achievability} \label{sec:achievability}
This section sequentially addresses the proofs of achievable coding rate and error probability and then computes their ratio to obtain a lower bound on $\alpha$.

\subsection{Coding Rate}

The following proposition is an extension of the Berger-Tung coding scheme \cite{Tung1978} for an arbitrary number of sensors and for including context information at the CEO. As we will show in the sequel, the Berger-Tung source coding attains the optimal error exponent asymptotically as $L\to\infty$. By specializing to the case $|\mathcal{S}|=1$, we note that the Berger-Tung source coding readily attains the optimal error exponent for the original CEO problem considered in~\cite{BergerZV1996}.

\begin{prop} \label{prop:coding_rate}
	Fix some distribution $\prod_{\ell=1}^L p_\ell(u_\ell|y_\ell)$ and function $\wh{x}(u^L, s)$ such that $\mathbb{E} [ d(X, \wh{X}) ] \le D$. Then, a rate tuple $(R_1, \ldots, R_L)$ is achievable if for every $\mathcal{L} \subset [1:L]$,
	\begin{align}
		R(\mathcal{L}):=\sum_{\ell \in \mathcal{L}} R_\ell > I(U(\mathcal{L}); Y^L | U(\mathcal{L}^c), S ). \label{eq:rate_tuple}
	\end{align}
	Moreover, for any $\delta > 0$, the choice of $R_\ell > I(U_\ell; Y_\ell | X, S) + \delta$ asymptotically satisfies \eqref{eq:rate_tuple} for every $\mathcal{L} \subset [1:L]$ as $L \to \infty$.
\end{prop}

\begin{IEEEproof}
	The proof is based on the Berger-Tung coding scheme \cite{Tung1978}, \cite[Chapter 12.1]{ElGamalK2011} with some adaptations that include the context information given to the CEO. 
	
	\medskip
	\noindent\emph{Codebook generation:} Let $\e > \e' > 0$ and fix a distribution $\prod_{\ell=1}^L p_\ell(u_\ell|y_\ell)$. For each $\ell \in [1:L]$, $m_\ell \in [1:2^{nR_\ell}]$, and $k_\ell \in [1:2^{n\Rh_\ell} ]$, randomly and independently generate sequences $u^n_\ell(m_\ell, k_\ell)$, each according to $\prod_{t=1}^n p_{U_\ell}(u_{\ell}(t))$, where $p_{U_\ell}(u) = \sum_{x, s, y} p(x, s) p(y|x, s) p_{\ell}(u|y)$. The codebook 
	\begin{align*}
		\mathcal{C}= \left\{ u_\ell^n(m_\ell, k_\ell): \ell \in [1:L], m_\ell \in [1:2^{nR_\ell}], k_\ell \in [1:2^{n\Rh_\ell}] \right\}
	\end{align*}
	is revealed to both the sensors (encoders) and the CEO (decoder).
	
	\medskip
	\noindent\emph{Encoding:} Upon observing $y_\ell^n$, encoder $\ell$ finds an index pair $(m_\ell, k_\ell)\in [1:2^{nR_\ell}] \times [1:2^{n\Rh_\ell}]$ such that
	\begin{align*}
		(u^n_\ell(m_\ell, k_\ell), y^n_\ell) \in \aepvar,
	\end{align*}
	where $\aepvar$ stands for the typical set \cite{CoverT1991, ElGamalK2011}. If there is more than one such index pair, the encoder selects one of them uniformly at random. If there is no such index pair, encoder $\ell$ selects an index pair from $[1:2^{nR_\ell}] \times [1:2^{n\Rh_\ell}]$ uniformly at random. Encoder $\ell$ sends the index $m_\ell$.
	
	\medskip
	\noindent\emph{Decoding:} Upon observing the index tuple $(m_1,\ldots, m_L)$ sent from the encoders and context information $s^n$, the decoder finds the unique index tuple $(\kh_1,\ldots, \kh_L)\in[1:2^{n\Rh_1}] \times \cdots \times [1:2^{n\Rh_L}]$ such that
	\begin{align*}
		(u^n_1(m_1, \kh_1), \dots, u^n_1(m_L, \kh_L), s^n) \in \aep.
	\end{align*}
	If there is no such index tuple, the decoder selects an index tuple from $[1:2^{n\Rh_1}] \times \cdots \times [1:2^{n\Rh_L}]$ at random.

	\medskip
	\noindent\emph{Error analysis:} Let $(M_\ell, K_\ell)$ be the pair of indices chosen from encoder $\ell$ and $\Kh_\ell$, $\ell\in[1:L]$ be the decoded indices. Define the error event
	\begin{align*}
		\Ec = \{ (U^n_1(M_1,\Kh_1),\ldots, U^n_L(M_L,\Kh_L), S^{n}, Y^{nL}, X^n) \not\in \aep\}
	\end{align*}
	and consider the following events:
	\begin{align*}
		&\Ec_1 = \{(U^n_\ell(M_\ell, K_\ell), Y^n_\ell) \not \in \aepvar \text{ for some } K_\ell \in [1:2^{n\Rh_\ell}], \ell \in[1:L]\}, \\
		&\Ec_2 = \{ (U^n_1(M_1, K_1), \ldots, U^n_L(M_L, K_L), S^{n}, Y^{nL}, X^n) \not\in \aep\}, \\
		&\Ec_3 = \{ (U^n_1(M_1, \kh_1), \ldots, U^n_L(M_L, \kh_L), S^{n}) \in \aep, \text{ for some } (\kh_1,\ldots, \kh_L)\neq (K_1,\ldots, K_L)\}.
	\end{align*}
	Then, by the union bound, 
	\begin{align*}
		\mathbb{P}(\Ec) \le \mathbb{P}(\Ec_1) + \mathbb{P}(\Ec_1^c \cap \Ec_2) + \mathbb{P}(\Ec_3).
	\end{align*}
	
	We bound each term. By the covering lemma \cite{ElGamalK2011}, $\mathbb{P}(\Ec_1)$ tends to zero as $n \to \infty$ if 
	\begin{align} \label{eq:code_cond1}
		R_\ell + \Rh_\ell > I(U_\ell; Y_\ell) + \d_1(\e'), \quad \ell\in[1:L].
	\end{align}
	By the Markov lemma \cite{ElGamalK2011}, $\mathbb{P}(\Ec_1^c\cap \Ec_2)$ tends to zero as $n\to\infty$. 
	
	By the packing lemma and similar steps as in \cite{MineroLK2015}, $\mathbb{P}(\Ec_3)$ tends to zero as $n \to \infty$ if
	\begin{align} \label{eq:code_cond2}
		\sum_{\ell \in \mathcal{L}} \Rh_\ell &< I(U(\mathcal{L}); U(\mathcal{L}^c), S) + \sum_{\ell\in\mathcal{L}}H(U_\ell) - H(U(\mathcal{L}))+\d_2(\e), ~~ \mathcal{L} \subset [1:L].
	\end{align}
	
	By eliminating $\Rh_\ell$ in \eqref{eq:code_cond1} and \eqref{eq:code_cond2}, we can show that the coding error tends to zero as $n\to\infty$ if
	\begin{align}
		R(\mathcal{L}) &>  \sum_{\ell\in\mathcal{L}} I(U_\ell; Y_\ell) -I(U(\mathcal{L}); U(\mathcal{L}^c), S) - \sum_{\ell\in\mathcal{L}} H(U_\ell) + H(U(\mathcal{L})) + \d' \nonumber \\ 
		&= H(U(\mathcal{L})|U(\mathcal{L}^c), S) - \sum_{\ell\in\mathcal{L}} H(U_\ell|Y_\ell) + \d' \nonumber \\
		&\stackrel{(a)}{=} H(U(\mathcal{L}) | U(\mathcal{L}^c), S) - \sum_{\ell\in\mathcal{L}} H(U_\ell | U(\mathcal{L} \cap [1:\ell-1]), Y^L, U(\mathcal{L}^c), S) + \d' \nonumber \\
		&= H(U(\mathcal{L})|U(\mathcal{L}^c), S) - H(U(\mathcal{L})| Y^L, U(\mathcal{L}^c), S) + \d' \nonumber \\
		&= I(U(\mathcal{L}); Y^L | U(\mathcal{L}^c), S ) + \d', \label{eq:BT_rate_region}
	\end{align}
	where (a) follows since the Markov chain $(Y(\ell^c), U(\ell^c), S) - Y_\ell - U_\ell$ holds. Thus, the coding strategy established above has a vanishing probability of coding error as $n \to \infty$ if the rate tuple satisfies \eqref{eq:BT_rate_region}. As $\delta'$ can be made arbitrarily small, it proves the first claim.
	
	Consider the second claim. To show this, rewrite \eqref{eq:BT_rate_region} as follows.
	\begin{align} \label{eq:achievability2}
		R(\mathcal{L}) &> I(U(\mathcal{L}); Y^L | U(\mathcal{L}^c), S ) + \delta' \nonumber \\
		&= \left( \sum_{\ell\in\mathcal{L}} I(U_\ell; Y_\ell| X, S) \right) + I(U(\mathcal{L}); X | U(\mathcal{L}^c), S ) + \delta',
	\end{align}
	the proof of which is given in Appendix. We choose our operating rate at
	\begin{align}
		R_\ell = I(U_\ell; Y_\ell | X, S) + \delta. \label{eq:rate}
	\end{align}
	In the following, we show that the rate \eqref{eq:rate} asymptotically satisfies \eqref{eq:achievability2} as $L \to \infty$, i.e., we show that for any $\delta > 0$,
	\begin{align*}
		\left( \sum_{\ell\in\mathcal{L}} I(U_\ell; Y_\ell| X, S) \right) + |\mathcal{L}|\delta > \left( \sum_{\ell\in\mathcal{L}} I(U_\ell; Y_\ell| X, S) \right) + I(U(\mathcal{L}); X | U(\mathcal{L}^c, S) ).
	\end{align*}
	when $L$ is sufficiently large. Hence, it suffices to prove
	\begin{align*}
		\frac{1}{|\mathcal{L}|} I(U(\mathcal{L}); X | U(\mathcal{L}^c), S ) < \delta
	\end{align*}
	when $L$ is sufficiently large.
	
	To this end, we divide the problem into two cases, namely, $|\mathcal{L}|>\frac{L}{2}$ and $|\mathcal{L}|\le\frac{L}{2}$. The former is trivial since 
	\begin{align*}
		I(U(\mathcal{L}); X | U(\mathcal{L}^c), S ) \le H(X)
	\end{align*}
	where the upper bound is independent of $L$. For the latter, 
	\begin{align*}
		I(U(\mathcal{L}); X | U(\mathcal{L}^c), S ) \le H(X| U(\mathcal{L}^c), S).
	\end{align*}	
	Let $\wt{X}(\mathcal{L}^c) := f(U(\mathcal{L}^c), S)$ be a reconstruction function for identifying $X$.
	By Fano's inequality, we have
	\begin{align*}
		H(X| U(\mathcal{L}^c), S) \le H_2(P^{(|\mathcal{L}^c|)}_e) + P^{(|\mathcal{L}^c|)}_e \log(|\Xc|-1)
	\end{align*}
	where $P^{(|\mathcal{L}^c|)}_e = \mathbb{P} [\wt{X}(\mathcal{L}^c) \neq X]$. The upper bound vanishes as $|\mathcal{L}^c| \to \infty$, see Section \ref{subsec:detection_error} and Proposition \ref{prop:error_freq}, which are essentially the analysis of ML detection with many observations \cite{Poor1988, MoulinV2019}. It proves the second claim.	
\end{IEEEproof}

The following corollary is an immediate consequence of Proposition \ref{prop:coding_rate}, which will be used in Section \ref{subsec:final_achievability} to compute a lower bound on $\alpha$.
\begin{cor} \label{cor:sum_rate_bound}
	Fix pmfs $p(j)$, $\{p(u | y, j)\}_j$, and a function $\wh{x}(u^L, s)$ such that $\mathbb{E} [ d(X, \wh{X}) ] \le D$. Then, if $\mathcal{L} = [1:L]$, then for any $\delta, \delta' > 0$, there exists a code such that
	\begin{align*}
		\Rsum < L \cdot I(U;Y|X,S,J) + \delta,
	\end{align*}
	and the coding error is less than $\delta'$ if $L$ is sufficiently large.
\end{cor}
\begin{IEEEproof}
	Let $\mathcal{L} = [1:L]$ and take an auxiliary random variable $J \in \mathcal{J} = [1:L]$ such that $\mathbb{P}[U=u|Y,J=j] = \mathbb{P}[U_j=u|Y]$. Then,
	\begin{align*}
		\sum_{\ell \in [1:L]} I(U_\ell; Y_\ell | X, S) &= L \cdot \frac{1}{L} \sum_{\ell \in [1:L]} I(U_\ell; Y_\ell | X, S) \\
		&= L \sum_{j \in \mathcal{J}} p(j) I(U; Y | X, S, J=j) \\
		&= L \cdot I(U;Y|X,S,J).
	\end{align*}
	Then, we can take $n$ and $L$ sufficiently large so that $\delta, \delta'$ requirements in the claim are satisfied.
\end{IEEEproof}

\subsection{Detection Error Probability} \label{subsec:detection_error}
Note that the final probability of detection error is
\begin{align*}
	P_e \le \mathbb{P}[ \mathcal{E} ] + (1-\mathbb{P}[ \mathcal{E} ]) \cdot \frac{1}{n} \mathbb{E}[d_H(X^n, \wh{X}^n) | \mathcal{E}^c].
\end{align*}
Since limits $n \to \infty$ and $L, \Rsum \to \infty$ will be taken sequentially, $\mathbb{P}[ \mathcal{E} ] \to 0$ provided that the coding rate tuple satisfies \eqref{eq:rate_tuple}.\footnote{Alternatively, one may take a joint limit witn $n$ growing sufficiently faster than $L, \Rsum$ so that $\mathbb{P}[ \mathcal{E} ] \to 0$.} Hence, it is sufficient to focus only on $\frac{1}{n} \mathbb{E}[d_H(X^n, \wh{X}^n) | \mathcal{E}^c ]$ term. The following proposition bounds this term.

\begin{prop} \label{prop:error_freq}
	Assuming the previous coding scheme,
	\begin{align*}
		\frac{1}{n} \mathbb{E}[d_H(X^n, \wh{X}^n) | \mathcal{E}^c ] \le \exp \left( -L \max_{(s, x_1, x_2)} \max_{\lambda \in [0,1]} \mathbb{E}_{J} \left[ d_{\lambda}(p_{x_1, s, J}(u), p_{x_2, s, J}(u)) \right] \right).
	\end{align*}
\end{prop}

\begin{IEEEproof}
	Suppose that the CEO successfully recovers $\wh{U}_\ell^{n}$ for all $\ell$. In other words, the CEO has $S^n$ and $\wh{U}^{nL} = U^{nL}$, from which it attempts to detect $X^n$. The CEO will detect $X(t)$ for all $t$ in a symbol-wise manner using the same detection rule; then, its error probability for each symbol should be the same since for all $t$, $U_\ell(1), \ldots, U_\ell(n)$ are i.i.d.~generated from a fixed distribution. It in turn implies that $\frac{1}{n} \mathbb{E}[d_H(X^n, \wh{X}^n) | \mathcal{E}^c ] = \mathbb{P}[\wh{X}(t) \ne X(t) | \wh{U}^{nL} = U^{nL}]$ for any $t$. Therefore, it suffices to consider an error probability at an arbitrary time $t$. For notational brevity, let $P$ be the probability measure conditioned on $\wh{U}^{nL} = U^{nL}$, that is, $P[\cdot] = \mathbb{P}[ \cdot |\wh{U}^{nL} = U^{nL}]$.
	
	Notice that $P[\wh{X}(t) \ne X(t)] = \mathbb{P}[\wh{X}(t) \ne X(t) | \wh{U}^{nL} = U^{nL}]$ can be represented as follows.
	\begin{align*}
		P[\wh{X}(t) \ne X(t)] &= \sum_s p(s) \sum_{x_1} p(x_1|s) P[ \wh{X}(t) \ne x_1 |S(t)=s ] \\
		&= \sum_s p(s) \sum_{x_1 \ne x_2} p(x_1|s) P[ \wh{X}(t) = x_2 | X(t) = x_1, S(t)=s ].
	\end{align*}
	Further, if we rewrite $P[ \wh{X}(t) = x_2 | X(t) = x_1, S(t)=s ]$ in an exponential form, that is,
	\begin{align*}
		P[ \wh{X}(t) = x_2 | X(t) = x_1, S(t)=s ] = \exp(-L \eta(x_2|x_1,s) + \text{const})
	\end{align*}
	where $\eta(x_2|x_1,s)$ is an exponent possibly dependent on $L$, then the total error probability is represented and bounded as
	\begin{align*}
		P[\wh{X}(t) \ne X(t)] &= \sum_{x_1 \ne x_2, s} p(x_1, s) \exp(-L \eta(x_2|x_1,s) + \text{const}) \\
		&\le \sum_{x_1 \ne x_2, s} \exp(-L \eta(x_2|x_1,s) + \text{const}).
	\end{align*}
	One can first make an observation that the prefactor $p(x_1, s)$ does not change the exponent of the upper bound, which in turn implies that assuming any $p(x,s)$ does not change the exponent of the upper bound. The second observation is that the CEO needs to maximize the smallest $\eta(x_2|x_1,s)$, as it dominates the entire $P[\wh{X}(t) \ne X(t)]$ among all triplets of $(s, x_1, x_2)$ for $\eta(x_2|x_1, s)$.
	
	Having $s$ fixed and assuming $p(x|s)$ is uniform since assuming any $p(x,s)$ does not change the optimal exponent, the problem of interest at time $t$ now becomes the $|\mathcal{X}|$-ary hypothesis testing problem with observations $u^{nL}$. Hence it is optimal to perform the maximum likelihood (ML) detection. As the testing problem is statistically identical across time due to its i.i.d.~coding scheme, we omit the temporal index $t$ for brevity. Let $\mathcal{U}_{x}^L = \mathcal{U}_{x}^L(s)$ be the set of vectors such that
	\begin{align*}
		\mathcal{U}_{x}^L(s) := \left\{ u^L: p( u^L | x, s ) \ge p( u^L | x', s ) ~~ \forall x' \ne x \right\},
	\end{align*}
	that is, the ML decision region for $\wh{X}=x$. Also, note that for any $\lambda \in [0,1]$,
	\begin{align*}
		\mathcal{U}_{x}^L &= \left\{ u^L: p( u^L | x, s ) \ge p( u^L | x', s ) ~~ \forall x' \ne x \right\} \\
		&= \left\{ u^L: p^{1-\lambda}( u^L | x', s ) p^{\lambda}( u^L | x, s ) \ge p( u^L | x', s ) ~~ \forall x' \ne x \right\}.
	\end{align*}
	Using this notation, we can represent error probabilities. First,
	\begin{align*}
		P[\wh{X}=x_2 | X=x_1, s] &= \sum_{ u^L \in \mathcal{U}_{x_2}^L} p( u^L | x_1, s ) \\
		&\stackrel{(a)}{\le} \sum_{ u^L \in \mathcal{U}_{x_2}^L} p^{1-\lambda}( u^L | x_1, s ) p^{\lambda}( u^L | x_2, s ) \\
		&\stackrel{(b)}{\le} \sum_{ u^L \in \mathcal{U}^L } p^{1-\lambda}( u^L | x_1, s ) p^{\lambda}( u^L | x_2, s ).
	\end{align*}
	where (a) follows from the definition of $\mathcal{U}_{x_2}^L$ and (b) follows since $\mathcal{U}_{x_2}^L \subset \mathcal{U}^L $. Repeating the same argument, but using $\lambda \leftarrow 1-\lambda$ instead,
	\begin{align*}
		P[\wh{X}=x_1|X=x_2, s] &\le \sum_{ u^L \in \mathcal{U}^L } p^{1-\lambda}( u^L | x_1, s ) p^{\lambda}( u^L | x_2, s ).
	\end{align*}
	Therefore, we have
	\begin{align*}
		&P[\wh{X}=x_2 | X=x_1, s] + P[\wh{X}=x_1|X=x_2, s] \\
		&\le 2 \sum_{u^L \in \mathcal{U}^L } p^{1-\lambda}( u^L | x_1, s ) p^{\lambda}( u^L | x_2, s ) \\
		&= 2 \prod_{\ell=1}^L \sum_{u_\ell } p_{\ell}^{1-\lambda}( u_\ell | x_1, s ) p_{\ell}^{\lambda}( u_\ell |  x_2, s ) \\
		&= \exp \left( \sum_{\ell=1}^L \log \sum_{u_\ell} p_{\ell}^{1-\lambda}( u_\ell | x_1, s ) p_{\ell}^{\lambda}( u_\ell | x_2, s ) + o(L) \right).
	\end{align*}
	
	Taking an auxiliary random variable $J \in \mathcal{J} = [1:L]$ such that $\mathbb{P}[U=u|Y,J=j] = \mathbb{P}[U_j=u|Y]$ and noting that $p_\ell(u_\ell | x, s) = \sum_y p(y|x,s)p_\ell(u_\ell|y)$, we have
	\begin{align*}
		&P[\wh{X}=x_2|X=x_1, s] + P[\wh{X}=x_1|X=x_2, s] \\
		&\le \exp \bigg( L \cdot \frac{1}{L}\sum_{\ell=1}^L \log  \sum_{u_\ell} p_{\ell}^{1-\lambda}( u_\ell | x_1, s ) p_{\ell}^{\lambda}( u_\ell | x_2, s ) + o(L) \bigg) \\
		&= \exp \bigg( L \sum_{j} p(j) \log  \sum_{u} p^{1-\lambda}( u | x_1, s, j ) p^{\lambda}( u | x_2, s, j ) + o(L) \bigg) \\
		&= \exp \Bigg( -L \mathbb{E} \left[ -\log  \sum_{u} p^{1-\lambda}( u | x_1, s, J ) p^{\lambda}( u | x_2, s, J ) \right] + o(L) \Bigg).
	\end{align*}
	As $\lambda \in [0,1]$ is arbitrary, optimizing $\lambda$ gives
	\begin{align*}
		&P[\wh{X}=x_2|X=x_1, s] + P[\wh{X}=x_2|X=x_1, s] \\
		&\le \exp \Bigg( -L \max_{\lambda} \mathbb{E} \left[ -\log  \sum_{u} p^{1-\lambda}( u | x_1, s, J ) p^{\lambda}( u | x_2, s, J ) \right] + o(L) \Bigg).
	\end{align*}

	Recall that there are $\binom{|\mathcal{X}|}{2} |\mathcal{S}|$ triplets of $(s, x_1, x_2)$, which is finite and having any $p(x,s)$ does not change the error exponent. Using notation $\doteq$ that denotes the equality in the first order of exponent,
	\begin{align*}
		P[\wh{X} \ne X] &= \sum_s p(s) \sum_{x_1 \ne x_2} p(x_1|s) P[ \wh{X}(t) = x_2 | X(t) = x_1, S(t)=s ] \nonumber \\
		&\doteq \sum_s \frac{1}{|\mathcal{S}|} \sum_{x_1 \ne x_2} \frac{1}{|\mathcal{X}|} P[ \wh{X}(t) = x_2 | X(t) = x_1, S(t)=s ] \nonumber \\
		&\le \exp \left( -L \min_{(s, x_1, x_2)} \max_{\lambda \in [0,1]} \mathbb{E} \left[ -\log \sum_{u} p^{1-\lambda}( u |x_1, s, J ) p^{\lambda}( u |x_2, s, J ) \right] + o(L) \right).
	\end{align*}

	Since $\frac{1}{n} \mathbb{E}[d_H(X^n, \wh{X}^n) | \mathcal{E}^c] = P[\wh{X} \ne X] \doteq P_e$, we finally have
	\begin{align*}
		-\log P_e(\Rsum) &\ge L \min_{(s, x_1, x_2)} \max_{\lambda \in [0,1]} \mathbb{E} \left[ -\log \sum_{u} p^{1-\lambda}( u | x_1, s, J ) p^{\lambda}( u | x_2, s, J ) \right] + o(L) \\
		&= L \min_{(s, x_1, x_2)} \max_{\lambda \in [0,1]} \mathbb{E} \left[ d_{\lambda}(p_{x_1, s, J}(u), p_{x_2, s, J}(u)) \right] + o(L).
	\end{align*}
	where $p_{x, s, j}(u) := p_{U|X,S,J}(u|x,s,j)$.
\end{IEEEproof}

\subsection{Final Step of Achievability} \label{subsec:final_achievability}
Combining Corollary \ref{cor:sum_rate_bound} and Proposition \ref{prop:error_freq}, we have the following lower bound on $\alpha$ at average individual rate $c := I(U;Y|X,S,J)$.
{\small
\begin{align*}
	\alpha( c ) &\ge \lim_{L \to \infty} \frac{L \min_{(s, x_1, x_2)} \max_{\lambda \in [0,1]} \mathbb{E}_{J} \left[ d_{\lambda}(p_{x_1, s, J}, p_{x_2, s, J}) \right] + o(L)}{L \cdot I(U;Y|X,S,J) + o(L)} \\
	&= \frac{ \min_{(s, x_1, x_2)} \max_{\lambda \in [0,1]} \mathbb{E}_{J} \left[ d_{\lambda}(p_{x_1, s, J}, p_{x_2, s, J}) \right] }{I(U;Y|X,S,J)}.
\end{align*}
}
Optimizing the lower bound over all possible $p(j)$ and test channel $p(u|y,j)$ at average rate $c$ and sending $c$ that stands for average individual rate to zero prove the expression of $\alpha$ in the claim.

The cardinality bound on $\mathcal{U}$ is based on the support lemma \cite[Lemma 3.3.4]{CsiszarK1997}, \cite[Appendix C]{ElGamalK2011}. Suppose $p^*(u|j)$ and $p^*(u|y,j)$ taking values over a large alphabet $\mathcal{U}$ achieve the lower bound. The following argument shows that there exists $p(u|j), p(u|y,j)$ over a smaller subset $\mathcal{U}' \subset \mathcal{U}$ is sufficient to reproduce the lower bound.

Notice that the Markov chain $(X,S)-Y-U$ implies 
\begin{align*}
	I(U;Y|X,S,J) &= I(U;Y|J) - I(X,S;U|J) \\
	&= H(Y|J) - H(Y|U,J) - H(X,S|J) + H(X,S|U,J).
\end{align*}
Then, new $p(u|j), p(u|y,j)$ should satisfy the following constraints:
\begin{itemize}
	\item $p(y|j) = \sum_u p(u|j) p(y|u,j)$ for all $y, j$; then, it also preserves $H(Y|j), H(X,S|j)$ for all $j$. This gives $(|\mathcal{Y}|-1) |\mathcal{J}|$ constraints,
	
	\item $H(Y|U,j) - H(X,S|U,j) = \sum_{u} p(u|j) (H(Y|U=u, j) - H(X,S|U=u,j))$ for all $j$; this gives $|\mathcal{J}|$ constraints.
	
	\item Also, our new $p(u|j)$ should preserve
	\begin{align*}
		d_{\lambda^*}(p_{x_1, s, J=j}, p_{x_2, s, J=j}) = d_{\lambda^*} \left( \sum_y p(y|x_1,s)p^*(u|y,j), \sum_y p(y|x_2,s)p^*(u|y,j) \right)
	\end{align*}
	for all $(s, x_1, x_2)$ and $j$, where $\lambda^* = \lambda^*(s, x_1, x_2)$ is the maximizer. Hence, our new $p(u|j), p(u|y,j)$ should also preserve $\binom{|\mathcal{X}|}{2} |\mathcal{S}| |\mathcal{J}|$ Chernoff divergences, which are continuous in $p(u|y,j)$.
\end{itemize}

Therefore, the number of constraints is in total
\begin{align*}
	\left( \binom{|\mathcal{X}|}{2} |\mathcal{S}| + |\mathcal{Y}| \right) |\mathcal{J}|.
\end{align*}
Then, by the support lemma \cite[Lemma 3.3.4]{CsiszarK1997}, \cite[Appendix C]{ElGamalK2011}, $|\mathcal{U}| \le \left( \binom{|\mathcal{X}|}{2} |\mathcal{S}| + |\mathcal{Y}| \right) |\mathcal{J}|$.

We defer the proof of the cardinality bound on $\mathcal{J}$ to Section \ref{sec:cardinality} as it is new in the literature and tighter than the support lemma.

\section{Proof of Converse} \label{sec:converse}
Like the achievability proof, this section sequentially addresses the proofs of converse coding rate and error probability and then computes their ratio to obtain an upper bound on $\alpha$.

\subsection{Coding Rate Analysis}
First, recall the notation that we will use throughout the proof.
\begin{align*}
	X(t^c) &= (X(1), \ldots, X(t-1), X(t+1), \ldots, X(n)), \\
	S(t^c) &= (S(1), \ldots, S(t-1), S(t+1), \ldots, S(n)).
\end{align*}
Then, we can derive the following lower bound on the individual rate.
\begin{align*}
	R_\ell &= \frac{1}{n} \log | \mathcal{C}_\ell^{(n)} | \\
	&\ge \frac{1}{n} H(C_\ell) \ge \frac{1}{n} H(C_\ell | X^n, S^n) \\
	&\ge \frac{1}{n} H(C_\ell | X^n, S^n)  - \frac{1}{n} H(C_\ell | X^n, Y_\ell^n, S^n) \\
	&= \frac{1}{n} I(C_\ell; Y_\ell^n | X^n, S^n) \\
	&= \frac{1}{n} \left( H(Y_\ell^n | X^n, S^n) - H(Y_\ell^n | X^n, S^n, C_\ell) \right) \\
	&= \frac{1}{n} \sum_{t=1}^n \bigg( H(Y_\ell(t) | X(t), X(t^c), S(t), S(t^c), Y_{\ell}^{t-1}) - H(Y_\ell(t) | X(t), X(t^c), S(t), S(t^c), Y_{\ell}^{t-1}, C_\ell) \bigg) \\
	&\ge \frac{1}{n} \sum_{t=1}^n \bigg( H(Y_\ell(t) | X(t), X(t^c), S(t), S(t^c)) - H(Y_\ell(t) | X(t), X(t^c), S(t), S(t^c), C_\ell) \bigg),
\end{align*}
where the last inequality follows from two facts that 1) the Markov chain $Y_\ell(t) - (X^n, S^n) - Y_\ell^{t-1}$ holds and 2) removing conditions increases entropy. 

To represent the codeword $C_\ell$, we take a random variable $U_\ell(t, x(t^c), s(t^c))$ such that for given $x(t^c)$ and $s(t^c)$, its joint distribution with $X(t), Y_\ell(t), S(t)$ is
\begin{align}
	&\mathbb{P}[ Y_\ell(t) = y, U_\ell(t, x(t^c), s(t^c)) = c | X(t) = x, S(t) = s ] \nonumber \\
	&= p(y|x) \mathbb{P}[ C_\ell = c | Y_\ell(t) = y, S(t) = s, X(t) = x, X(t^c) = x(t^c), S(t^c) = s(t^c) ] \nonumber \\
	&= p(y|x) \mathbb{P}[ C_\ell = c | Y_\ell(t) = y, X(t^c) = x(t^c), S(t^c) = s(t^c) ]. \label{eq:u_defi_converse}
\end{align}
In other words, the Markov chain $(X(t), S(t)) - Y_\ell(t) - U_\ell(t, x(t^c), s(t^c))$ holds if $x(t^c), s(t^c)$ are given. Then, the lower bound on $R_\ell$ can be rewritten using $U_\ell$ and expectation over $X(t^c), S(t^c)$ as follows.
\begin{align}
	R_\ell \ge \frac{1}{n} \sum_{t=1}^n \mathbb{E}_{X(t^c), S(t^c)} \left[ I( Y_\ell(t); U_\ell(t, X(t^c), S(t^c)) | X(t), S(t) ) \right]. \label{eq:individual_rate_LB}
\end{align}

\subsection{Detection Error Probability}
Note that the error probability of our interest at time $t$ can be decomposed as follows.
\begin{align*}
	\mathbb{P} [ \wh{X}(t) \ne X(t) ] &= \sum_{x_1, s} p(x_1, s) \mathbb{P}[\wh{X}(t) \ne x | X(t) = x_1, S(t) = s] \\
	&\ge \sum_{x_1, s} p(x_1, s) \sum_{x_2 \ne x_1} \mathbb{P}[\wh{X}(t) = x_2 | X(t) = x_1, S(t) = s] \\
	&\ge \min_{x_1, s} p(x_1, s) \cdot \max_{(s, x_1, x_2)} \mathbb{P}[\wh{X}(t) = x_2 | X(t) = x_1, S(t) = s] \\
	&\ge \text{const} \cdot \max_{(s, x_1, x_2)} \mathbb{P}[\wh{X}(t) = x_2 | X(t) = x_1, S(t) = s].
\end{align*}
The converse in this subsection mainly focuses on showing that each $\mathbb{P}[\wh{X}(t) = x_2 | X(t) = x_1, S(t) = s]$ exponentially decays with $L$. Since the probability of error is a linear sum of individual errors corresponding to all triplets $(s, x_1, x_2)$, finding the triplet yielding the largest $\mathbb{P}[\wh{X}(t) = x_2 | X(t) = x_1, S(t) = s]$, or equivalently, the smallest individual error exponent, gives the greatest lower bound on $\mathbb{P} [ \wh{X}(t) \ne X(t) ]$.

We will derive a genie-aided lower bound on the detection error probability. Without loss of generality, suppose that $X(t) = x_1$ is the true value. Recalling that the CEO has $s(t)$ (and $s(t^c)$ as well), suppose that a genie chooses $x_2 \in \mathcal{X} \setminus \{x_1\}$ uniformly at random and provides information for the CEO that $X(t)$ belongs to $\{x_1, x_2\}$ together with $X(t^c) = x(t^c)$. Note that a pair $(x_1, x_2)$ will be chosen with probability $\frac{p(x_1|s) + p(x_2|s)}{|\mathcal{X}|-1}$.

As $x_1$ and $x_2$ are given candidates for $X(t)$, it is optimal for the CEO to perform the binary hypothesis testing using received codewords and given information $x(t^c), s^n$. In particular, fix $X(t^c) = x(t^c), S^n =s^n$ and focus only on obtaining a lower bound on $\mathbb{P}[\wh{X}(t) = x_2 | X(t) = x_1, S(t) = s]$ and $\mathbb{P}[\wh{X}(t) = x_1 | X(t) = x_2, S(t) = s]$ at time $t$. Since the effective prior probability is $\left( \frac{p(x_1|s)}{p(x_1|s) + p(x_2|s)}, \frac{p(x_2|s)}{p(x_1|s) + p(x_2|s)} \right)$, it is optimal to perform the log-likelihood ratio (LLR) test:
\begin{align*}
	\wh{X}(t) = \begin{cases}
		x_1 & \textrm{if } \LLR \left(c^L | x(t^c), s^n \right) \ge -\log \frac{p(x_1|s)}{p(x_2|s)}, \\
		x_2 & \textrm{if } \LLR \left(c^L | x(t^c), s^n \right) < -\log \frac{p(x_1|s)}{p(x_2|s)},
	\end{cases}
\end{align*}
where
\begin{align*}
	\LLR \left(c^L | x(t^c), s^n \right) &:= \log \frac{ p( c^L | X(t) = x_1, x(t^c), s^n ) }{ p( c^L | X(t) = x_2, x(t^c), s^n ) } \\
	&= \sum_{\ell=1}^L \log \frac{ p(c_\ell | X(t) = x_1, x(t^c), s^{n} ) }{ p( c_\ell | X(t) = x_2, x(t^c), s^{n} )} \\
	&=: \sum_{\ell=1}^L \LLR_\ell( c_\ell | x(t^c), s^{n} ),
\end{align*}
where the equality follows since codewords are independent conditioned on $x^n, s^n$.

Let $\mathcal{U}_{x_1}=\mathcal{U}_{x_1}(s)$ and $\mathcal{U}_{x_2}=\mathcal{U}_{x_2}(s)$ be the decoding regions corresponding to $x_1, x_2$, respectively:
\begin{align*}
	\mathcal{U}_{x_1}(s) &= \left\{ c^L : \LLR \left(c^L | x(t^c), s^n \right) \ge -\log \frac{p(x_1|s)}{p(x_2|s)} \right\}, \\
	\mathcal{U}_{x_2}(s) &= \left\{ c^L : \LLR \left(c^L | x(t^c), s^n \right) < -\log \frac{p(x_1|s)}{p(x_2|s)} \right\}.
\end{align*}
Let $P_{\lambda, C^L}(c^L), P_{\lambda, C_\ell} (c_\ell)$ respectively be the geometric mixture distributions with parameter $\lambda \in [0,1]$ as follows.
\begin{align*}
	P_{\lambda, C^L} (c^L) &:= \frac{ p^{1-\lambda} ( c^L | X(t) = x_1, x(t^c), s^n ) p^{\lambda} ( c^L | X(t) = x_2, x(t^c), s^n ) }{ \sum_{c^L} p^{1-\lambda} ( c^L | X(t) = x_1, x(t^c), s^n ) p^{\lambda} ( c^L | X(t) = x_2, x(t^c), s^n ) }, \\
	P_{\lambda, C_\ell} (c_\ell) &:= \frac{ p^{1-\lambda} ( c_\ell | X(t) = x_1, x(t^c), s^{n} ) p^{\lambda} (c_\ell | X(t) = x_2, x(t^c), s^{n} ) }{ \sum_{c_\ell} p^{1-\lambda} ( c_\ell | X(t) = x_1, x(t^c), s^{n} ) p^{\lambda} ( c_\ell | X(t) = x_2, x(t^c), s^{n} ) }.	
\end{align*}
Then, $P_{\lambda, C^L} (c^L) = \prod_\ell P_{\lambda, C_\ell} (c_\ell)$ holds due to independence. 

For $\epsilon > 0$, choose $\lambda = \lambda(\epsilon)$ such that
\begin{align}
	\mathbb{E}_{ P_{\lambda, C^L} } [ \LLR (C^L | x(t^c), s^n) ] = L \epsilon - \log \frac{p(x_1|s)}{p(x_2|s)}. \label{eq:our_lambda}
\end{align}
That is, the mean of $\LLR$ with respect to $P_{\lambda, C^L}$ is slightly greater than its decision threshold. Since $\LLR(C^L | x(t^c), s^n)$ is a sum of $L$ $\LLR_\ell$'s, from the Chebyshev inequality,
\begin{align*}
	&P_{\lambda, C^L} \bigg( c^L : \LLR (c^L | x(t^c), s^n) \ge - \log \frac{p(x_1|s)}{p(x_2|s)} \bigg) \\
	&\ge P_{\lambda, C^L} \bigg( c^L : \Big|  \LLR (c^L | X(t^c), S^n) - \mathbb{E}_{P_{\lambda, C^L}} [ \LLR (C^L | x(t^c), s^n) ] \Big| < L\epsilon  \bigg) \\
	&\ge 1 - \frac{ \sum_{\ell=1}^L \sigma_\ell^2 }{ L^2 \epsilon^2 },
\end{align*}
where $\sigma_\ell^2$ is the variance of $\LLR_\ell (C_\ell | x(t^c), s^n)$ with respect to $P_{\lambda, C_\ell}$. Therefore, we know that $\lim_{L \to \infty} P_{\lambda, C^L}(\mathcal{U}_{x_1}) = 1$ as $L \to \infty$.

Upon receiving codewords, the CEO determines (i.e., processes) a binary event whether $\wh{X}=x_1$ or $\wh{X}=x_2$. By the data processing inequality of the KL divergence,
\begin{align*}
	&D\left( P_{\lambda, C^L}(c^L) \| p(c^L | X(t) = x_2, x(t^c), s^n) \right) \\
	&\ge D\left( P_{\lambda, C^L}(\mathcal{U}_{x_2}) \| p(\mathcal{U}_{x_2} | X(t) = x_2, x(t^c), s^n) \right) \\
	&= P_{\lambda, C^L}(\mathcal{U}_{x_1}) \log \frac{ P_{\lambda, C^L}(\mathcal{U}_{x_1}) }{ p( \mathcal{U}_{x_1} | X(t) = x_2, x(t^c), s^n ) } \\
	&~~~~ ~~~~ ~~~~ + P_{\lambda, C^L}(\mathcal{U}_{x_2}) \log \frac{ P_{\lambda, C^L}(\mathcal{U}_{x_2}) }{ p( \mathcal{U}_{x_2} | X(t) = x_2, x(t^c), s^n) } \\
	&= -H_2( P_{\lambda, C^L}(\mathcal{U}_{x_1}) ) - P_{\lambda, C^L}(\mathcal{U}_{x_1}) \log p( \mathcal{U}_{x_1} | X(t) = x_2, x(t^c), s^n ) \\
	&~~~~ ~~~~ ~~~~ - P_{\lambda, C^L}(\mathcal{U}_{x_2}) \log p( \mathcal{U}_{x_2} | X(t) = x_2, x(t^c), s^n ) \\
	&\ge -H_2( P_{\lambda, C^L}(\mathcal{U}_{x_1}) ) \\
	&~~~~ ~~~~ ~~~~ - P_{\lambda, C^L}(\mathcal{U}_{x_1}) \log p( \mathcal{U}_{x_1} | X(t) = x_2, x(t^c), s^n ),
\end{align*}
where $H_2$ is the binary entropy and the last inequality holds since $P_{\lambda, C^L} (\mathcal{U}_{x_2}) \log p(\cdots ) \le 0$. Rearranging terms, we have
\begin{align*}
	&\log p( \mathcal{U}_{x_1} | X(t) = x_2, x(t^c), s^n ) \\
	&\ge \frac{1}{P_{\lambda, C^L}(\mathcal{U}_{x_1})} \left( -D\big( P_{\lambda, C^L}(\cdot) \| p( \cdot | X(t) = x_2, x(t^c), s^n) \big) -H_2( P_{\lambda, C^L}(\mathcal{U}_{x_1})) \right),
\end{align*}
which in turn implies
\begin{align*}
	p( \mathcal{U}_{x_1} | X(t) = x_2, x(t^c), s^n ) \ge e^{ -D\left( P_{\lambda, C^L} \| p(\cdot | X(t) = x_2, x(t^c), s^n) \right) + o(L) }.
\end{align*}
In addition, since $P_{\lambda, C^L} (c^L) = \prod_\ell P_{\lambda, C_\ell} (c_\ell)$ holds, the additive property of the KL divergence yields
\begin{align*}
	p( \mathcal{U}_{x_1} | X(t) = x_2, x(t^c), s^n ) \ge e^{ - \sum_{\ell} D\left( P_{\lambda, C_\ell} \| p( \cdot | X(t) = x_2, x(t^c), s^n) \right) + o(L) }.
\end{align*}
Similarly, we also have
\begin{align*}
	p( \mathcal{U}_{x_2} | X(t) = x_1, x(t^c), s^n ) \ge e^{ -\sum_{\ell} D\left( P_{\lambda, C_\ell} \| p( \cdot | X(t) = x_1, x(t^c), s^n) \right) + o(L) }.
\end{align*}

We define $\beta_{x_1}, \beta_{x_2}$ respectively as the expectations of the above probabilities, or in other words, expected Type I and Type II error probabilities respectively, which will be used soon.
\begin{align}
	\beta_{x_1}(t;s) &:= \mathbb{E}_{ X(t^c), S(t^c) } \left[ p( \mathcal{U}_{x_2} | X(t) = x_1, X(t^c), S(t)=s, S(t^c) ) \right], \nonumber \\
	\beta_{x_2}(t;s) &:= \mathbb{E}_{ X(t^c), S(t^c) } \left[ p( \mathcal{U}_{x_1} | X(t) = x_2, X(t^c), S(t)=s, S(t^c) ) \right]. \label{eq:beta_1_2}
\end{align}

\subsection{Final Step of Converse}
In this subsection, we will assume $\lambda^\circ$ satisfying
\begin{align*}
	\mathbb{E}_{ P_{\lambda^\circ, C^L} } [ \LLR (C^L | x(t^c), s^n) ] = 0,
\end{align*}
or equivalently,
\begin{align}
	D\left( P_{\lambda^\circ, C^L} \| p( \cdot | X(t) = x_1, x(t^c), s^n) \right) = D\left( P_{\lambda^\circ, C^L} \| p(\cdot | X(t) = x_2, x(t^c), s^n) \right), \label{eq:lambda_circ}
\end{align}
which is slightly different from that in \eqref{eq:our_lambda}. If we normalize both sides by $1/L$, the difference between using $\lambda^\circ$ and using $\lambda$ in \eqref{eq:our_lambda} is asymptotically negligible as in \cite[p.~899]{BergerZV1996}. Hence, we will use $\lambda^\circ$ in the sequel without additional arguments.

Consider the ratio of our interest. Combining the individual rate lower bound \eqref{eq:individual_rate_LB} and the genie-aided lower bound on the error probabilities \eqref{eq:beta_1_2}, we have the following with $\lambda$ being chosen so that \eqref{eq:lambda_circ} holds.
\begin{align*}
	&\frac{-\log P_e^{(n)}}{\Rsum} \le \frac{ -\log \frac{1}{n} \sum_{t=1}^n \mathbb{P}[X(t) \ne \wh{X}(t)] }{ \frac{1}{n} \sum_{\ell=1}^L \sum_{t=1}^n \mathbb{E}_{X(t^c), S(t^c)} [ I( Y_\ell(t); U_\ell(t) | X(t), S(t) ) ] } \nonumber \\
	&\le \frac{ -\log \frac{1}{n} \sum_{t=1}^n \sum_{s(t)} p(s(t)) \sum_{x_1 \ne x_2}  \left( \frac{p(x_1|s(t))\beta_{x_1}(t;s(t)) + p(x_2|s(t))\beta_{x_2}(t;s(t)) }{|\mathcal{X}|-1}  \right) }{ \frac{1}{n} \sum_{\ell=1}^L \sum_{t=1}^n \mathbb{E}_{X(t^c), S(t^c)} [ I( Y_\ell(t); U_\ell(t) | X(t), S(t) ) ] } \nonumber \\
	&\le \frac{ -\log \frac{1}{n} \sum_{t=1}^n \sum_{s(t), x_1 \ne x_2} \frac{\text{const}}{|\mathcal{X}|-1} \mathbb{E}_{X(t^c), S(t^c)} \left[ e^{ -\sum_{\ell} D\left( P_{\lambda, C_\ell} \| p( \cdot | X(t) = x_1, X(t^c), s(t), S(t^c)) \right) + o(L) } \right] }{ \frac{1}{n} \sum_{\ell=1}^L \sum_{t=1}^n \mathbb{E}_{X(t^c), S(t^c)} [ I( Y_\ell(t); U_\ell(t) | X(t), S(t) ) ] }.
\end{align*}

Next, we interchange $\mathbb{E}_{X(t^c), S(t^c)}$ and $\sum_{s(t), x_1 \ne x_2} \frac{\text{const}}{|\mathcal{X}|-1}$ and then apply Jensen's inequality twice on the numerator. As $\frac{\text{const}}{|\mathcal{X}|-1}$ is a finite constant, it can be captured by $o(L)$, which gives 
\begin{align*}
	\frac{-\log P_e^{(n)}}{\Rsum} &\le \frac{ -\log \frac{1}{n} \sum_{t=1}^n \sum_{s(t), x_1 \ne x_2} \frac{\text{const}}{|\mathcal{X}|-1} \mathbb{E}_{X(t^c), S(t^c)} \left[ e^{ -\sum_{\ell} D\left( P_{\lambda, C_\ell} \| p( \cdot | X(t) = x_1, X(t^c), s(t), S(t^c)) \right) + o(L) } \right] }{ \frac{1}{n} \sum_{\ell=1}^L \sum_{t=1}^n \mathbb{E}_{X(t^c), S(t^c)} [ I( Y_\ell(t); U_\ell(t) | X(t), S(t) ) ] } \\
	&\le \frac{ \frac{1}{n} \sum_{t=1}^n -\log \mathbb{E}_{X(t^c), S(t^c)} \left[  \sum_{s(t), x_1 \ne x_2} e^{ -\sum_{\ell} D\left( P_{\lambda, C_\ell} \| p( \cdot | X(t) = x_1, X(t^c), s(t), S(t^c)) \right) + o(L) } \right] }{ \frac{1}{n} \sum_{\ell=1}^L \sum_{t=1}^n \mathbb{E}_{X(t^c), S(t^c)} [ I( Y_\ell(t); U_\ell(t) | X(t), S(t) ) ] } \\
	&\le \frac{ \frac{1}{n} \sum_{t=1}^n \mathbb{E}_{X(t^c), S(t^c)} \left[ -\log \sum_{s(t), x_1 \ne x_2} e^{ -\sum_{\ell} D\left( P_{\lambda, C_\ell} \| p( \cdot | X(t) = x_1, X(t^c), s(t), S(t^c)) \right) + o(L) } \right] }{ \frac{1}{n} \sum_{\ell=1}^L \sum_{t=1}^n \mathbb{E}_{X(t^c), S(t^c)} [ I( Y_\ell(t); U_\ell(t) | X(t), S(t) ) ] } \\
	&\le \frac{ \frac{1}{n} \sum_{t=1}^n \mathbb{E}_{X(t^c), S(t^c)} \left[ -\log \max_{s(t), x_1, x_2} e^{ -\sum_{\ell} D\left( P_{\lambda, C_\ell} \| p( \cdot | X(t) = x_1, X(t^c), s(t), S(t^c)) \right) + o(L) } \right] }{ \frac{1}{n} \sum_{\ell=1}^L \sum_{t=1}^n \mathbb{E}_{X(t^c), S(t^c)} [ I( Y_\ell(t); U_\ell(t) | X(t), S(t) ) ] } \\
	&= \frac{ \frac{1}{n} \sum_{t=1}^n \mathbb{E}_{X(t^c), S(t^c)} \left[ \min_{s(t), x_1, x_2} \sum_{\ell} D\left( P_{\lambda, C_\ell} \| p( \cdot | X(t) = x_1, X(t^c), s(t), S(t^c)) \right) + o(L) \right] }{ \frac{1}{n} \sum_{\ell=1}^L \sum_{t=1}^n \mathbb{E}_{X(t^c), S(t^c)} [ I( Y_\ell(t); U_\ell(t) | X(t), S(t) ) ] } \\
	&= \frac{ \sum_{t=1}^n \mathbb{E}_{X(t^c), S(t^c)} \left[ \min_{s(t), x_1, x_2} \sum_\ell D\left( P_{\lambda, C_\ell} \| p( \cdot | X(t) = x_1, X(t^c), s(t), S(t^c)) \right) + o(L) \right] }{ \sum_{t=1}^n \sum_{\ell=1}^L \mathbb{E}_{X(t^c), S(t^c)} [ I( Y_\ell(t); U_\ell(t) | X(t), S(t) ) ] }.
\end{align*}

Note that $\frac{\sum_t A_t}{\sum_t B_t} \le \max_t \frac{A_t}{B_t}$ holds for positive $A_t$ and $B_t$, which leads
\begin{align*}
	\frac{-\log P_e^{(n)}}{\Rsum} \le \max_{ t, x(t^c), s(t^c) } \frac{ \min_{s(t), x_1, x_2} \sum_{\ell=1}^L D\left( P_{\lambda, C_\ell} \| p( \cdot | X(t) = x_1, x(t^c), s(t), s(t^c)) \right) + o(L) }{  \sum_{\ell=1}^L I( Y_\ell(t); U_\ell(t) | X(t), S(t) ) }.
\end{align*}

From our definition of $U_\ell$ in \eqref{eq:u_defi_converse} and the Markov chain $(X(t), S(t)) - Y_\ell(t) - U_\ell(t, X(t^c), S(t^c))$, for any coding and estimation scheme and for any $\epsilon > 0$, if $L$ is sufficiently large,
\begin{align*}
	\frac{-\log P_e^{(n)}}{\Rsum} \le \max_{U_\ell: (X,S) - Y_\ell - U_\ell} \frac{ \min_{s, x_1, x_2} \frac{1}{L} \sum_{\ell=1}^L D\left( P_{\lambda, C_\ell}(u_\ell) \| p( u_\ell | x_1, s) \right) }{  \frac{1}{L} \sum_{\ell=1}^L I( Y_\ell; U_\ell | X, S ) } + \epsilon.
\end{align*}
Also, to represent the averages of quantities across sensors into a single term, we introduce an auxiliary random variable $J \in \mathcal{J} := [1:L]$ that is independent of $X, Y_\ell$ and satisfies $(X,S)-(Y,J)-U$ where $U$ is a super random variable defined on the union of $\mathcal{U}_\ell$ with $\mathbb{P}[U=u|Y, J=j] = \mathbb{P}[U_j=u|Y]$. Then,
\begin{align*}
	\frac{-\log P_e^{(n)}}{\Rsum} &\le \max_{p(j), p(u|y,j)} \frac{ \min_{s, x_1, x_2} \mathbb{E}_{J} \left[ D\left( P_{\lambda, C, J}(u) \| p(u|x_1,s,J) \right) \right] }{ I( Y; U | X, S, J ) } + \epsilon \\
	&= \max_{p(j), p(u|y,j)} \frac{ \min_{s, x_1, x_2} \max_{\lambda} \mathbb{E}_{J} \left[ d_\lambda (p_{x_1,s,J}, p_{x_1,s,J}) \right] }{ I( Y; U | X, S, J ) } + \epsilon,
\end{align*}
where the equality follows since at our choice of $\lambda$ \eqref{eq:lambda_circ},
\begin{align*}
	\mathbb{E} \left[ D\left( P_{\lambda, C, J}(u) \| p(u|x_1,s,J) \right) \right] = \mathbb{E} \left[ D\left( P_{\lambda, C, J}(u) \| p(u|x_2,s,J) \right) \right] = \max_\lambda \mathbb{E} \left[ d_\lambda (p_{x_1,s,J}, p_{x_2,s,J}) \right]
\end{align*}
holds \cite{MoulinV2019}.

Sending $L \to \infty$, we have a bound
\begin{align*}
	\frac{-\log P_e^{(n)}}{\Rsum} \le \max_{p(j), p(u|y,j)} \frac{ \min_{(s, x_1, x_2)} \max_{\lambda} \mathbb{E}_{J} \left[ d_\lambda (p_{x_1,s,J}, p_{x_2,s,J}) \right] }{ I( Y; U | X, S, J ) }.
\end{align*}
Finally, sending the mutual information to zero completes the proof. Note that $\mathcal{J}$ has countably infinite elements in its current form; the refinement is given in the next section.

\section{Number of Groups $|\mathcal{J}|$} \label{sec:cardinality}
This section particularly proves the following statement on the number of groups required to achieve $\alpha$. Also, extensions of our proof are discussed.

\begin{thm}[Restatement of Theorem \ref{thm:alpha} on $|\mathcal{J}|$] \label{thm:reduced_cardinality}
	Equation \eqref{eq:alpha} in Theorem \ref{thm:alpha} can be attained using $|\mathcal{J}|=\binom{|\mathcal{X}|}{2} |\mathcal{S}|$. That is, dividing sensors into at most $\binom{|\mathcal{X}|}{2} |\mathcal{S}|$ groups and all sensors in each group adopting the same compression code are asymptotically optimal.
\end{thm}

The proof relies on the following lemma on the solution to a linear fractional program (LFP). Note that the proof technique in \cite{Tsitsiklis1988} uses linear programs, which cannot be used for the CEO problem as \eqref{eq:alpha} is not linear in $p(j)$.

\begin{lem} \label{lem:LFP}
	For positive constants $a_{ij}, b_{ij}, c_1, c_2$ where $i \in [1:i_\maxx]$ and $j \in [1:j_\maxx]$ with $j_{\maxx} > i_{\maxx}$, the following is a linear fractional program of $j_{\maxx}$ variables $\{w_j\}_{j=1}^{j_{\maxx}}$:
	\begin{align*}
		\max_{w_j}\min_{i=1,\ldots,i_{\maxx}} ~~~ &\frac{ \sum_{j=1}^{j_{\maxx}} w_j a_{ij} + c_1 }{ \sum_{j=1}^{j_{\maxx}} w_j b_{ij} + c_2 } \\
		\text{subject to} ~~~ &w_j \ge 0 ~~ \forall j \in [1:j_\maxx] ~~ \text{ and } ~~ \sum_{j=1}^{j_{\maxx}} w_j=1.
	\end{align*}
	Let $\{w_j^*\}_{j=1}^{j_{\maxx}}$ be the optimal solution to the LFP. Then, $\{w_j^*\}_{j=1}^{j_{\maxx}}$ has at least $(j_{\maxx}-i_{\maxx})$ zeros, i.e., at most $i_{\maxx}$ variables could be nonzero.
\end{lem}

\begin{IEEEproof}[Proof of Lemma \ref{lem:LFP}]
	Note that the program can be converted into an equivalent LFP by introducing an additional variable $\gamma \ge 0$:
	\begin{align}
		\max ~~ &\gamma \label{eq:equiv_LP} \\
		\text{subject to} ~~~ &\frac{\sum_{j=1}^{j_{\maxx}} w_j a_{ij} + c_1}{\sum_{j=1}^{j_{\maxx}} w_j b_{ij} + c_2} \ge \gamma ~~~ \forall i \in [1:i_\maxx] \nonumber \\
		&w_j \ge 0 ~~~ \forall j \in [1:j_\maxx] ~~ \text{ and } ~~ \sum_{j=1}^{j_{\maxx}} w_j=1. \nonumber
	\end{align}
	The new program as well as the original LFP is a nonlinear program as  $\frac{\sum_{j=1}^{j_{\maxx}} w_j a_{ij} + c_1}{\sum_{j=1}^{j_{\maxx}} w_j b_{ij} + c_2}$ is nonlinear in $w_j$. However, note that letting $\gamma^*$ be the largest attainable value of \eqref{eq:equiv_LP}, i.e., $\gamma^* := \max \gamma$, the constraints in \eqref{eq:equiv_LP} hold for any $\gamma \in [0, \gamma^*]$. Using this, \eqref{eq:equiv_LP} can be solved by a sequence of feasibility tests of linear programs (LPs), which is described as follows.
	\begin{enumerate}
		\item[1)] Take an arbitrary initial $\gamma^\circ > 0$ and a sufficiently small error tolerance $\epsilon > 0$.
		
		\item[2)] Check the feasibility of an LP (how to check is described below):
		\begin{align*}
			&\frac{\sum_{j=1}^{j_{\maxx}} w_j a_{ij} + c_1}{\sum_{j=1}^{j_{\maxx}} w_j b_{ij} + c_2} \ge \gamma^\circ ~~~ \forall i \in [1:i_\maxx] \\
			\text{subject to} ~~~ &w_j \ge 0 ~~ \forall j \in [1:j_\maxx] ~ \text{ and } ~ \sum_{j=1}^{j_{\maxx}} w_j=1.
		\end{align*}
		
		\item[3-1)] If the LP is feasible, i.e., a solution to the program with $\gamma^\circ$ exists, repeat Step 2) with $\gamma^\circ \leftarrow \gamma^\circ + \epsilon$ until the program becomes infeasible. If it becomes infeasible, terminate the process.
		
		\item[3-2)] If the LP is infeasible, repeat Step 2) with $\gamma^\circ \leftarrow \gamma^\circ - \epsilon$ until the program becomes feasible. If it becomes feasible, terminate the process.
	\end{enumerate}
	Then, the solution to the last feasible LP is the solution to the original LFP within the error tolerance $\epsilon$. The accuracy of the solution obtained here can be easily improved: For instance, instead of termination, repeat Steps 2) and 3) with $\epsilon \leftarrow \epsilon/2$ until the desired accuracy is met.
	
	The feasibility test in Step 2) is indeed equivalent to solving the following LP and testing whether $t^*$ is nonnegative or not.
	\begin{align*}
		t^* &:= \max ~ t ~~~~ \text{subject to} \\
		~~ &\sum_{j=1}^{j_{\maxx}}\left( w_j a_{ij} - \gamma^\circ w_j b_{ij} \right) + c_1 - \gamma^\circ c_2 \ge t ~~ \forall i \in [1:i_\maxx] \\
		&~ w_j \ge 0 ~~~ \forall j \in [1:j_\maxx] ~~ \text{ and } ~~ \sum_{j=1}^{j_{\maxx}} w_j=1.
	\end{align*}
	If $t^*$ is nonnegative, then the LP in Step 2) is feasible. 
	
	Note that each LP has $(j_\maxx + 1)$ variables and $(i_{\maxx}+j_{\maxx}+1)$ constraints. Then, feasible variables form a polyhedron consisting of a set of $j_{\maxx}$-dimensional faces due to the equality constraint. Also, note that the solution to a feasible LP exists only at a vertex of its polyhedron, e.g., by the simplex method \cite{HillierL2005}. Since the polyhedron's faces are $j_{\maxx}$-dimensional and vertices occur when at least $j_{\maxx}$ inequality constraints are satisfied with equality. That means, at most $i_{\maxx}$ constraints are satisfied with strict inequality, which in turn implies that at most $i_{\maxx}$ variables are nonzero in the set of solution variables.
	
	Since the solution to each LP in Step 2) has at most $i_{\maxx}$ nonzeros, the solution to the original LFP has at most $i_{\maxx}$ nonzeros as well. It completes the proof.
\end{IEEEproof}

Now we are ready to prove Theorem \ref{thm:reduced_cardinality}, the bound on the number of groups.

\begin{IEEEproof}[Proof of Theorem \ref{thm:reduced_cardinality}]
	Note that the value of $\alpha$ in \eqref{eq:alpha} is a limit value; it formally means that for any $\epsilon > 0$ one can find a good pair of $p_J, \{p(u|y,j)\}_j$ that achieves $\alpha - \epsilon$. Fix an arbitrary $\epsilon > 0$, and let $p^\circ_J, \{p^\circ(u|y,j)\}_j$ be a pair that achieves $\alpha - \epsilon$. Let $p_J^\#$ be the maximizer with respect to $\{p^\circ(u|y,j)\}_j$, i.e.,
	{\small
	\begin{align}
		p_J^\# = \argmax_{p_J} \frac{ \min_{ (s, x_1, x_2) } \max_{\lambda} \mathbb{E}_J[d_\lambda (p_{x_1, s, J}, p_{x_2, s, J})] }{ I(U;Y|X,S,J) } \Bigg\vert_{ \{p^\circ(u|y,j)\}_j }. \label{eq:cardi_argmax}
	\end{align}
	}
	Then $p_J^\#, \{p^\circ(u|y,j)\}_j$ will achieve a value equal to or greater than $\alpha-\epsilon$. Note that $p_J^\#$ is possibly supported on more than $\binom{|\mathcal{X}|}{2} |\mathcal{S}|$ elements. We view \eqref{eq:cardi_argmax} as a linear fractional program and then prove that the same value can be attained with $p_J^*$ that is supported on at most $\binom{|\mathcal{X}|}{2} |\mathcal{S}|$ elements.
	
	Fixing $\{p^\circ(u|y,j)\}_j$, we have
	\begin{align*}
		\alpha - \epsilon \le \alpha' := \max_{p_J} \frac{ \min_{ (s, x_1, x_2) } \max_{\lambda} \mathbb{E}_J[d_\lambda ( p_{x_1, s, J}, p_{x_2, s, J} )] }{ I(U;Y|X,S,J) },
	\end{align*}
	where $p_J$ that maximizes the right side is indeed $p_J^\#$ by definition. Let $\lambda^*$ be the maximizer that attains $\alpha'$. Then, it can be rewritten as follows.
	\begin{align}
		\alpha - \epsilon \le \alpha' = \max_{p_J} \min_{ s, x_1, x_2 } \frac{ \sum_j p_J(j) d_{\lambda^*} (p_{x_1, s, j}, p_{x_2, s, j}) }{ \sum_j p_J(j) I(U;Y|X,S,J=j) }. \label{eq:before_reduce}
	\end{align}
	Note that the minimum in \eqref{eq:before_reduce} picks the smallest term among $\binom{|\mathcal{X}|}{2} |\mathcal{S}|$ ones which we can reindex with $i \in [1: i_\maxx]$ where $i_\maxx = \binom{|\mathcal{X}|}{2} |\mathcal{S}|$. Also, letting
	\begin{align*}
		w_j &\leftarrow p_J(j), ~~~ a_{ij} \leftarrow d_{\lambda^*}(p_{x_1, s, j}, p_{x_2, s, j}) \\
		b_{ij} &\leftarrow I(U;Y|X,S,J=j) ~~~ \text{for all } i
	\end{align*}
	gives the following LFP:
	\begin{align*}
		\alpha-\epsilon \le \alpha' &= \max_{p_J} \min_{s, x_1, x_2} \frac{ \sum_j p_J(j) d_{\lambda^*} (p_{x_1, s, j}, p_{x_2, s, j}) }{ \sum_j p_J(j) I(U;Y|X,S,J=j) } \\
		&= \max_{ \{w_j\}_j } \min_{i=1,\ldots, i_\maxx} \frac{ \sum_{j=1}^{j_{\maxx}} w_j a_{ij} }{ \sum_{j=1}^{j_{\maxx}} w_j b_{ij} }.
	\end{align*}
	Then, Lemma \ref{lem:LFP} proves that the optimizer $\{w_j^*\} = \{p_J^*(j)\}$ has at most $i_\maxx = \binom{|\mathcal{X}|}{2} |\mathcal{S}|$ nonzeros. As the argument holds for arbitrary $\epsilon$, we can see that \eqref{eq:alpha} can be attained within arbitrary accuracy using $p_J(j)$ having at most $|\mathcal{J}|=\binom{|\mathcal{X}|}{2} |\mathcal{S}|$ nonzero elements as well.
\end{IEEEproof}

It is immediate that if $|\mathcal{S}|=1$, the problem reduces to the original CEO problem, and our result tightens the number of groups in \cite[Theorem 1]{BergerZV1996}. 
Also, note that such a form of linear fractional program is common in the CEO problems where the number of sensors and sum rate grow without bound. Thus, the proof can be used for other CEO problems as well. For instance, when there is no context information and the sensors are statistically nonexchangeable, which is the model for different types of sensors having distinct observational channels, our result improves the number of groups required to achieve the optimal exponent.
\begin{thm}[Theorem 2 in \cite{BergerZV1996} with bound on $|J \times K|$ improved]
	Let $J$ be an independent auxiliary random variable on space $\mathcal{J}$, and let $\mathcal{K}$ be the alphabet, over which $K$ indicates the type of sensors. Also, let $p(u|y,j,k)$ be any probability mass function on $\mathcal{U}$. Then,
	\begin{align*}
		\alpha = \lim_{c \to 0} \max_{ \substack{p_{J,K}, p_{U|Y,J,K}: \\ I(U;Y|X,K,J)=c } } \frac{ \min_{x_1, x_2} \max_{\lambda \in [0,1]} \mathbb{E}_{J,K} \left[ d_{\lambda}(p_{x_1, J, K}(u), p_{x_2, J, K}(u)) \right] }{I(U;Y|X,K,J)},
	\end{align*}
	where
	\begin{align*}
		p_{x, j, k}(u) := p_{U|X,J,K}(u|x,j,k)  = \sum_y p(y|x,k)p(u|y,j,k).
	\end{align*}
	Also, it is sufficient to have
	\begin{align*}
		|J \times K| = \binom{|\mathcal{X}|}{2}, ~~~ |\mathcal{U}| &= \left(\binom{|\mathcal{X}|}{2} + |\mathcal{Y}| \right) |J \times K| + 2.
	\end{align*}
\end{thm}
The result implies that the sensors' types and compression schemes are jointly partitioned into at most $\binom{|\mathcal{X}|}{2}$ groups, which reduces the bound $\binom{|\mathcal{X}|}{2} + 2$ in \cite[Theorem 2]{BergerZV1996}.

If all $b_{ij}$'s in Lemma \ref{lem:LFP} are the same, then the linear fractional program is indeed a linear program. Recalling that if the number of sensors tends to infinity, the error exponent of symbolwise compression problems is indeed equivalent to optimizing the minimax of expected Chernoff divergences \cite{Tsitsiklis1988, NitinawaratAV2013}. Therefore, Lemma \ref{lem:LFP} reproduces Tsitsiklis's number of groups $\binom{|\mathcal{X}|}{2}$ in \cite{Tsitsiklis1988}. Also, it concludes that if sensors' types could be different \cite{NitinawaratAV2013}, then the sensors' types and compression schemes should be jointly partitioned into at most $\binom{|\mathcal{X}|}{2}$ groups. This conclusion is missing in \cite{NitinawaratAV2013}.

\section{Example} \label{sec:example}
We consider a simple and tractable Gaussian example to demonstrate the effect of the availability of context information $S$. Suppose that the source is binary, $\mathcal{X} = \{0,1\}$, and each sensor observes the source via an additive white Gaussian noise (AWGN) channel, together with independent external Gaussian context information $S$ being added. That is, $Y_\ell = X + N_\ell + S$, where $N_\ell \sim \mathcal{N}(0, \sigma_N^2)$, $S \sim \mathcal{N}(0, \sigma_S^2)$, and $p(x,s) = p(x)p(s)$. The realization of $S$ is known to the CEO. This can be thought of as an abstract and analytically tractable model for object recognition: $X$ represents whether an object exists or not, and there are $L$ cameras taking the object's (single-pixel) pictures $\{Y_\ell\}_\ell$ corrupted by additive Gaussian noise $N_\ell$ and external lighting $S$. The fusion center is aware of the intensity of sunlight and thus knows $S$.

Then, with sufficiently fine quantization, Theorem \ref{thm:alpha} characterizes $\alpha$. However, the exact evaluation of it is challenging since it requires test channel optimization that is nonconvex. Instead, we use a Gaussian test channel for the sake of tractability, i.e., $U = Y + V$ where $V \sim \mathcal{N}(0, \sigma_V^2)$, which gives a closed-form lower bound on $\alpha$. As the codeword variable is $U=X+S+N+V$, the context information simply shifts the center of the Gaussian distribution of $U$. The Chernoff divergence is invariant under shift; thus, simply setting $S=0$, the minimum over $S$ in the numerator can be removed without any change of value. Also, $|\mathcal{J}|=1$ is sufficient since $\mathcal{X}$ is binary and $S$ is fixed. This gives the following lower bound expression.
\begin{align}
	\alpha \ge \lim_{c \to 0} \max_{ p_{U|Y}: I(U;Y|X,S)=c } \frac{ \max_{\lambda \in [0,1]} d_{\lambda}(p_{x_1=0, s=0}(u), p_{x_2=1, s=0}(u)) }{I(U;Y|X,S)},\label{eq:gaussian_evaluation}
\end{align}
where $p_{U|Y}$'s are Gaussian test channels.

\begin{figure}[t]
	\centering
	\includegraphics[width=3.0in]{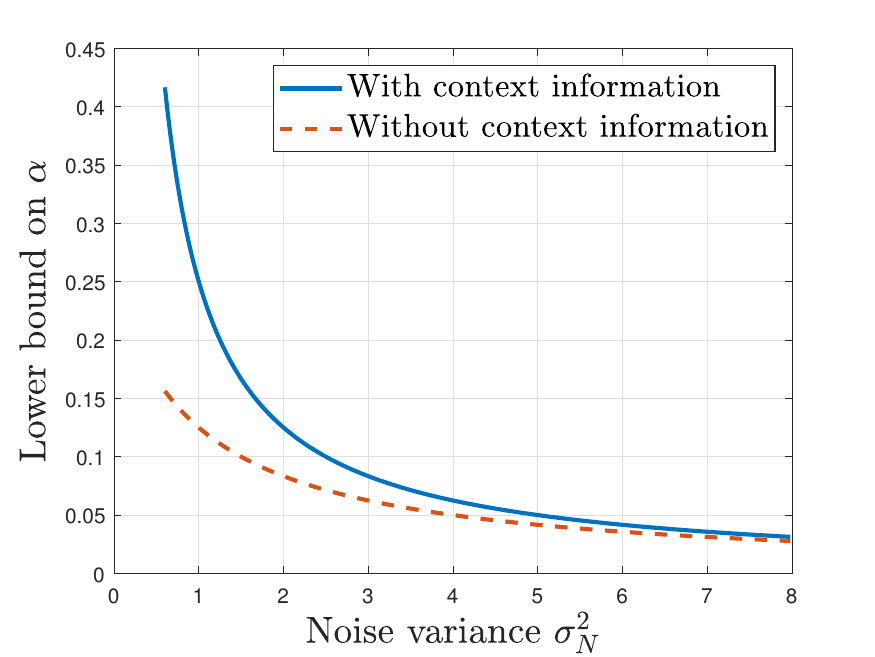}
	\caption{Numerical evaluation of the lower bound on $\alpha$. }
	\label{fig:eval}
\end{figure}

The denominator has a well-known closed-form expression,
\begin{align*}
	I(U;Y|X,S) &= h(U|X,S) - h(U|X,S,Y) \\
	&= h(N + V) - h(V) \\
	&= \frac{1}{2} \log \left( 1+ \frac{\sigma_N^2}{\sigma_V^2} \right).
\end{align*}
As we also know that $p_{X, S=0}(u) \sim \mathcal{N}(X, \sigma_N^2 + \sigma_V^2)$, the numerator can be optimized as follows.
\begin{align*}
	\lambda^* &= \argmax_{\lambda} d_\lambda(p_{0, 0}, p_{1, 0}) = \argmin_{\lambda} \int_{\mathcal{U}} p_{0, 0}^{1-\lambda}(u) p_{1, 0}^{\lambda}(u) du \\
	&= \argmin_{\lambda} \int_{\mathcal{U}} p_{0, 0}(u) \left( \frac{p_{1, 0}(u)}{p_{0, 0}(u)} \right)^{\lambda} du \\
	&= \argmin_{\lambda} \int_{\mathcal{U}} \frac{1}{\sqrt{2\pi (\sigma_N^2+\sigma_V^2)}} e^{- \frac{(u-\lambda)^2}{2(\sigma_N^2+\sigma_V^2)}} \cdot e^{\frac{\lambda^2 - \lambda}{2(\sigma_N^2 + \sigma_V^2)}} du \\
	&= \argmin_{\lambda} e^{\frac{\lambda^2 - \lambda}{2(\sigma_N^2 + \sigma_V^2)}} = \argmin_{\lambda} \lambda^2 - \lambda \\
	&= \frac{1}{2}.
\end{align*}
Therefore, the numerator is $d_{1/2} (p_{0,0}, p_{1,0}) = \frac{1}{8(\sigma_N^2 + \sigma_V^2)}$. Note that $\sigma_V^2$ is the only variable in the right side of \eqref{eq:gaussian_evaluation}. Since $I(U;Y|X,S) \to 0$ implies $\sigma_V^2 \to \infty$, \eqref{eq:gaussian_evaluation} is equivalent to
\begin{align*}
	\alpha \ge \lim_{\sigma_V^2 \to \infty} \frac{1}{4 (\sigma_N^2 + \sigma_V^2) \log \left( 1+ \frac{\sigma_N^2}{\sigma_V^2} \right)}.
\end{align*}
In this regime, the lower bound can be further simplified using the definition of constant $e$.
\begin{align*}
	\alpha &\ge \lim_{\sigma_V^2 \to \infty} \frac{1}{4 (\sigma_N^2 + \sigma_V^2) \log \left( 1+ \frac{\sigma_N^2}{\sigma_V^2} \right)} \\
	&= \lim_{\sigma_V^2 \to \infty} \frac{1}{4 (\sigma_N^2 + \sigma_V^2) \frac{\sigma_N^2}{\sigma_V^2} \log \left( 1+ \frac{\sigma_N^2}{\sigma_V^2} \right)^{\frac{\sigma_V^2}{\sigma_N^2}} } = \frac{1}{4 \sigma_N^2}.
\end{align*}
In contrast, if the context information is unavailable to the CEO, it is indeed the same as extra Gaussian noise from the CEO's perspective. Hence, without context information,
\begin{align*}
	\alpha \ge \frac{1}{4 (\sigma_N^2 + \sigma_S^2)}.
\end{align*}

Figure \ref{fig:eval} depicts the result for $\sigma_S^2=1$. One can observe that the achievable $\alpha$ is higher when context information is available. In particular, when the noise variance is small, the gain of context information is larger as $S$ dominates the compressed signal $U$. On the other hand, when the noise variance becomes larger, the gain vanishes since the independent noise dominates the compressed signal.

\section{Conclusion} \label{sec:conclusion}
This work investigates the CEO problem when the CEO is aware of context information. The asymptotically optimal error exponent per rate, as the number of sensors and sum rate tend to infinity is characterized. The proof extends the Berger-Tung coding scheme and the converse argument in \cite{BergerZV1996} taking into account context information. Also, it is proven based on the linear fractional programming (LFP) that having at most $\binom{|\mathcal{X}|}{2} |\mathcal{S}|$ groups is sufficient to achieve $\alpha$. If there is only a single letter for context information, i.e., $|\mathcal{S}|=1$, the result tightens Berger \textit{et al.}'s number of groups $\binom{|\mathcal{X}|}{2} + 2$ by $2$, and further, can be generalized to other CEO problems, e.g., the CEO problem with nonexchangeable sensors \cite[Theorem 2]{BergerZV1996}.

\appendix
\section{Omitted Part in Proof of Proposition \ref{prop:coding_rate}}

By the code construction, the Markov chain $X - Y^L - U(\mathcal{L})$ holds. Using it,
\begin{align*}
	I(U(\mathcal{L}); Y^L, X | U(\mathcal{L}^c), S ) &= I(U(\mathcal{L}); Y^L | U(\mathcal{L}^c), S ) + I(U(\mathcal{L}); X | Y^L, U(\mathcal{L}^c), S ) \\
	&= I(U(\mathcal{L}); Y^L | U(\mathcal{L}^c), S )
\end{align*}
Then, \eqref{eq:achievability2} can be shown as follows.
\begin{align*}
	I(U(\mathcal{L}); Y^L | U(\mathcal{L}^c), S ) &= I(U(\mathcal{L}); Y^L, X | U(\mathcal{L}^c), S ) \\
	&= I(U(\mathcal{L}); X | U(\mathcal{L}^c), S ) + I(U(\mathcal{L}); Y^L | X, U(\mathcal{L}^c), S ) \\
	&\stackrel{(a)}{=} I(U(\mathcal{L}); X | U(\mathcal{L}^c), S ) + I(U(\mathcal{L}); Y(\mathcal{L}) | X, U(\mathcal{L}^c), S ) \\
	&= I(U(\mathcal{L}); X | U(\mathcal{L}^c), S ) + H(U(\mathcal{L})| X, U(\mathcal{L}^c), S ) - H(U(\mathcal{L}) | X, U(\mathcal{L}^c), Y(\mathcal{L}), S ) \\
	&\stackrel{(b)}{=} I(U(\mathcal{L}); X | U(\mathcal{L}^c), S ) + H(U(\mathcal{L})| X, U(\mathcal{L}^c), S ) - H(U(\mathcal{L}) |X, Y(\mathcal{L}), S ) \\
	&= I(U(\mathcal{L}); X | U(\mathcal{L}^c), S ) + H(U(\mathcal{L})| X, U(\mathcal{L}^c), S ) \\
	&~~ - H(U(\mathcal{L})| X, S) + H(U(\mathcal{L})| X, S) - H(U(\mathcal{L}) |X, Y(\mathcal{L}), S )\\
	&= I(U(\mathcal{L}); X | U(\mathcal{L}^c), S ) +I(U(\mathcal{L}); Y(\mathcal{L})| X, S) -I(U(\mathcal{L}); U(\mathcal{L}^c)| X, S ) \\
	&\stackrel{(c)}{=} I(U(\mathcal{L}); X | U(\mathcal{L}^c), S ) +I(U(\mathcal{L}); Y(\mathcal{L})| X, S),
\end{align*}
where (a) follows from the Markov chain $Y(\mathcal{L}^c) - Y(\mathcal{L}) - U(\mathcal{L})$, (b) follows from the Markov chain $U(\mathcal{L}^c) - Y(\mathcal{L}) - U(\mathcal{L})$, and (c) follows from the Markov chain $U(\mathcal{L}) - (X,S) - U(\mathcal{L}^c)$. Further, the second term can be rewritten as follows.
\begin{align*}
	I(U(\mathcal{L}); Y(\mathcal{L})| X, S) &= H(Y(\mathcal{L})|X, S) - H(Y(\mathcal{L})|X, U(\mathcal{L}), S)\\
	&= \sum_{\ell\in\mathcal{L}} \left(H(Y_\ell|X, S, Y(\mathcal{L}_{\ell-1})) - H(Y_\ell|X, U(\mathcal{L}), S, Y(\mathcal{L}_{\ell-1}))\right) \\
	&= \sum_{\ell\in\mathcal{L}} \left(H(Y_\ell|X, S) - H(Y_\ell|X, U_\ell, S)\right)\\
	&= \sum_{\ell\in\mathcal{L}} I(U_\ell; Y_\ell|X, S)
\end{align*}
where $\mathcal{L}_{\ell-1} := \mathcal{L} \cap [1:\ell-1]$ and the Markov property $Y_\ell - (X, S) - (Y(\ell^c), U(\ell^c))$ is used. Therefore, we finally have
\begin{align*}
	I(U(\mathcal{L}); Y^L | U(\mathcal{L}^c), S ) &= \left( \sum_{\ell\in\mathcal{L}} I(U_\ell; Y_\ell| X, S) \right) + I(U(\mathcal{L}); X | U(\mathcal{L}^c), S ).
\end{align*}

\end{document}